\documentclass{aa}
\usepackage{graphicx}
\usepackage{txfonts}
\usepackage[draft]{hyperref}
\usepackage{natbib,twoopt}
\hypersetup{
  colorlinks=true,   %% links colored instead of frames
  urlcolor=blue,     %% external hyperlinks
  linkcolor=blue     %% internal latex links (eg Fig)
}
\bibpunct{(}{)}{;}{a}{}{,} %% natbib format for A&A and ApJ

\makeatletter
\newcommand{\bibnote}[2]{\global\@namedef{#1note}{#2}}
\newcommand{\biblink}[2]{\global\@namedef{#1link}{#2}}
\makeatother

\makeatletter
  \protected\def\stonyslink{%
     \def\hyper@linkstart##1##2{}\let\hyper@linkend\@empty}
  \newcommandtwoopt{\citeads}[3][][]{%
   \href{http://adsabs.harvard.edu/abs/#3}%
        {\stonyslink \citealp[#1][#2]{#3}}%   %% Rutten, 2000
   \biblink{#3}{\href{http://adsabs.harvard.edu/abs/#3}{ADS}}}
 \newcommandtwoopt{\citepads}[3][][]{%
   \href{http://adsabs.harvard.edu/abs/#3}%
        {\stonyslink \citep[#1][#2]{#3}}%     %% (Rutten 2000)
   \biblink{#3}{\href{http://adsabs.harvard.edu/abs/#3}{ADS}}}
 \newcommandtwoopt{\citetads}[3][][]{%
   \href{http://adsabs.harvard.edu/abs/#3}%
        {\stonyslink \citet[#1][#2]{#3}}%     %% Rutten (2000)
  \biblink{#3}{\href{http://adsabs.harvard.edu/abs/#3}{ADS}}}
 \newcommandtwoopt{\citeyearads}[3][][]{%
   \href{http://adsabs.harvard.edu/abs/#3}%
        {\stonyslink \citeyear[#1][#2]{#3}}%  %% 2000
   \biblink{#3}{\href{http://adsabs.harvard.edu/abs/#3}{ADS}}}
\makeatother

\begin{document}

   \title{A systematic survey of the dynamics of Uranus Trojans}

   \author{Lei Zhou\inst{1,2} \and Li-Yong Zhou\inst{1,2,3} \and Rudolf Dvorak\inst{4} \and Jian Li\inst{1,2,3}}
	\authorrunning{Zhou et al}
	\offprints{Zhou L.-Y., \email zhouly@nju.edu.cn}

	\institute{School of Astronomy and Space Science, Nanjing University, 163 Xianlin Avenue, Nanjing 210046, China 
	\and Key Laboratory of Modern Astronomy and Astrophysics in Ministry of Education, Nanjing University, Nanjing 210046, China 
	\and Institute of Space Astronomy and Extraterrestrial Exploration (NJU \& CAST)
	\and Universit\"{a}tssternwarte Wien, T\"{u}rkenschanzstr. 17, 1180 Wien, Austria}

   \date{}

% \abstract{}{}{}{}{} 
% 5 {} token are mandatory
 
  \abstract
   {The discovered Uranus Trojan (UT hereafter) $2011\ {\rm QF_{99}}$, as well as several candidates, has been reported to be on unstable orbits. This implies that the stability region around the triangular Lagrange points $L_4$ and $L_5$ of Uranus should be very limited.}
   {In this paper, we aim to locate the stability region for UTs and find out the dynamical mechanisms responsible for the structures in the phase space. The null detection of primordial UTs also needs to be explained.}
   {Using the spectral number as the stability indicator, we construct the dynamical maps on the $(a_0,i_0)$ plane. The proper frequencies of UTs are determined precisely with a frequency analysis method so that we can depict the resonance web via a semi-analytical method. We simulate the radial migration by introducing an artificial force acting on planets to mimic the capture of UTs.}%k with different time scales. (1 and 10\,Myr).}
   {Two main stability regions are found, one each for the low-inclination ($0^\circ$--$14^\circ$) and high-inclination regime ($32^\circ$--$59^\circ$). There is also an instability strip in each of them, at $9^\circ$ and $51^\circ$ respectively. They are supposed to be related with $g-2g_5+g_7=0$ and $\nu_8$ secular resonances. All stability regions are in the tadpole regime and no stable horseshoe orbits exist for UTs. The lack of moderate-inclined UTs are caused by the $\nu_5$ and $\nu_7$ secular resonances, which could excite the eccentricity of orbits. The fine structures in the dynamical maps are shaped by high-degree secular resonances and secondary resonances. %The nodal secular resonances $\nu_{17}$ and $\nu_{18}$ are also involved. It is strange that 
   Surprisingly, the libration center of UTs changes with the initial inclination, and we prove it is related to the quasi 1:2 mean motion resonance (MMR) between Uranus and Neptune. However, this quasi resonance has ignorable influence on the long-term stability of UTs in the current planetary configuration. About 36.3\% and 0.4\% of the pre-formed orbits survive the fast and slow migrations (with migrating time scales of 1 and 10\,Myr) respectively, most of which are in high inclination. Since the low-inclined UTs are more likely to survive the age of the solar system, they make up 77\% of all such long-life orbits by the end of the migration, making a total fraction up to $4.06\times10^{-3}$ and $9.07\times10^{-5}$ of the original population for the fast and slow migrations, respectively. The chaotic capture, just like the depletion, results from the secondary resonances when Uranus and Neptune cross their mutual MMRs. However, the captured orbits are too hot to survive till today.}
   {{About 3.81\% UTs are able to survive the age of the solar system, among which 95.5\% are on low-inclined orbits with $i_0<7.5^\circ$. However,} the depletion of the planetary migration seems to prevent a large fraction of such orbits, especially for the slow migration model. {Based on the widely-adopted migration models, a swarm of UTs at the beginning of the smooth outward migration is expected and a fast migration is in favour if any primordial UTs are detected}.}


   \maketitle
%
%-----------------------------------------------------------------

\section{Introduction}\label{sec:intro}

Trojan asteroids are a set of celestial bodies sharing the orbit with their parent planet. They reside around two triangular Lagrange equilibrium points $L_4$ and $L_5$ with their orbits locked in the 1:1 mean motion resonance (MMR) with the parent planet, forming two swarms leading and trailing the planet by $60^\circ$ respectively. Although the triangular Lagrange points are dynamically stable for all planets of the solar system in the circular restricted three-body model \citepads[see e.g.][]{1999ssd..book.....M}, the perturbations from other planets make the dynamical behaviour of Trojans complicated in the realistic model. Since the first discovery of a Jupiter Trojan (588 Achilles) in 1906 \citepads{1907AN....174...47W}, more than 7000 Trojans have been detected around both $L_4$ and $L_5$ of Jupiter. At the same time, there have been 22 Neptune Trojans confirmed according to the Minor Planet Center\footnote{\url{https://minorplanetcenter.net}} (MPC) database. Actually, the Trojan cloud of Neptune is believed to be the second largest reservoir of asteroids larger than 50\,km in radius in the solar system, only after the Kuiper belt \citepads{2005ApJ...628..520C,2006Sci...313..511S,2010ApJ...723L.233S}. To the contrary, the only discovery of Uranus Trojan (UT hereafter) $2011\ {\rm QF_{99}}$ around $L_4$ \citepads{2013MPEC....F...19A} listed in the MPC database indicates a smaller stability window for Trojans of Uranus. As a matter of fact, \citetads{2002Icar..160..271N} has shown that the Trojan region of Uranus are mostly unstable although a few stability windows may exist \citepads{2010CeMDA.107...51D}. Compared with other giant planets in the solar system, Uranus {has} the smallest stability region according to \citetads{2003A&A...410..725M}. 
     
$2011\ {\rm QF_{99}}$ (for orbital elements, see e.g. \textit{Asteroids-Dynamic Site}\footnote{\url{http://hamilton.dm.unipi.it/astdys/}}, AstDyS-2) is deemed to be a Centaur temporarily captured into the Trojan region recently \citepads{2013Sci...341..994A}. In fact, Centaurs could be a significant supply of temporary UTs \citepads{2006MNRAS.367L..20H}. To be specific, \citetads{2013Sci...341..994A} found that at any time 0.4\% of the $a<34\,{\rm AU}$ Centaur population supplied from scattering trans-Neptunian objects (TNOs) is in co-orbital motion with Uranus.
   
In addition, $2014\ {\rm YX_{49}}$ is identified as the second UT, temporarily on a high-inclined ($i=25.5^\circ$) tadpole orbit around $L_4$ \citepads{2017MNRAS.467.1561D}. Moreover, it was reported that there exist several possible Uranus companions such as 83982 (Crantor), $2002\ {\rm VG_{131}}$, $2010\ {\rm EU_{65}}$ and $2015\ {\rm DB_{216}}$, which are supposed to be on the transient horseshoe, quasi-satellite or even recurring co-orbital orbits \citepads{2006Icar..184...29G,2013A&A...551A.114D,2014MNRAS.441.2280D,2015MNRAS.453.1288D}. A detailed analysis of the dynamical evolution of these objects show that ephemeral multi-body MMRs could trigger the capture, ejection and switching between resonant co-orbital states of transient Uranus companions. Meanwhile, Jupiter and Neptune are found to render the long-term stability of them \citepads{2013A&A...551A.114D,2014MNRAS.441.2280D,2015MNRAS.453.1288D,2017MNRAS.467.1561D}. Actually, the specific mechanisms behind these objects are different from each other, indicating complicated dynamical properties of Uranus co-orbital regions.
   
For a more general investigation, \citetads{2010CeMDA.107...51D} explored the stability regions of UTs for a range of inclinations up to $60^\circ$. They found four stability windows of initial inclinations in the ranges of $(0^\circ,7^\circ)$, $(9^\circ,13^\circ)$, $(31^\circ,36^\circ)$ and $(38^\circ,50^\circ)$, where the most stable orbits could survive the age of the solar system. Counterintuitively, high-inclined UTs may be more stable than those with low inclinations even though some instability strips caused by secular resonances raise in the high-inclination regime. In addition, the apsidal secular resonances with Jupiter and Uranus are proven to be responsible for the instability gap around $i_0=20^\circ$. Although some similar mechanisms had been identified earlier by \citetads{2003A&A...410..725M}, they claimed that there are no stable UTs with low libration amplitudes due to indirect perturbations from other planets. 
   
The observational surveys of UTs are not only challenging because of their far distance from the Sun, but also discouraged by their low survival fractions over the age of the solar system (about 1\% as predicted by previous theoretical investigations). For an initial density resembling the present Trojans of Jupiter, the present population of 10-km radii UTs was estimated to be less than a few tens \citepads{2002Icar..160..271N}.
      
The early dynamical evolution of the outer solar system is believed to be involved with the planetary migration \citepads{1984Icar...58..109F,1995AJ....110..420M,1999AJ....117.3041H}, which could play an important role in the depletion of primordial UTs. And, to evaluate the influence of planetary migration on the orbital behaviour of UTs may in turn assess the proposed scenarios of the early evolution of the solar system. During the migration, the possible 1:2 MMR crossing between Jupiter and Saturn could develop great instability on their pre-formed Trojans, and even empty the co-orbital region \citepads{1998AJ....116.2590G,2001AJ....122.3485M,2005Natur.435..462M,2009AJ....137.5003N}. For Neptune Trojans, it turns out that only 5\% of an initial population could survive the early stage of the solar system if a slow radial migration is taken into consideration \citepads{2004Icar..167..347K}. Inferred from the observed high-inclined orbits, the present population of jovian Trojans and Neptune Trojans is probably the result of the chaotic capture from a primordial disk during the migration \citepads{2005Natur.435..462M,2009AJ....137.5003N,2009MNRAS.398.1715L,2010MNRAS.405.1375L,2015Icar..247..112P}. The influence of planetary migration on potential UTs deserves an investigation. 
   
This paper provides detailed dynamical maps of UTs on the $(a_0,i_0)$ plane where the long-term dynamical properties can be reflected with a relatively short integration time. A thorough investigation of the resonance mechanisms is needed to explain the complex structure of the phase space. Although orbits surviving the age of the solar system have been proven to be present \citepads{2010CeMDA.107...51D}, a comprehensive search for the stability regions in the phase space where possible primordial UTs reside is still necessary for the sake of %because it is instructive for the 
observations. The answer to the question why the primordial UTs have not been detected since they exist is in urgent need. The formation and evolution of the early solar system may provide an explanation, and in turn, the detection or null detection of UTs from subsequent observations can also restrict the theory of the early stage of the solar system.
   
The remainder of this paper is organized as follows. In Sect.~2, we briefly introduce the dynamical model alongside with the methods used for further analysis. In Sect.~3, we present the dynamical map on the $(a_0,i_0)$ plane and determine the stability regions. Furthermore, we try to explain the excitation of the orbits as well as the displacement of the libration centers. The resonance web is depicted in Sect.~4 with the help of a frequency analysis method. In virtue of a long time integration of orbits, Sect.~5 shows the long-term stability of UTs. We then discuss the influence of planetary migration on the depletion of primordial UTs in Sect.~6. Finally, Sect.~7 presents the conclusions and discussions.
   
%-----------------------------------------------------------------

\section{Model and Method}\label{sec:modmtd}
	
In this paper, we evaluate the orbital stability of UTs via numerical simulations and spectral analysis. Then we try to locate the resonances responsible for the structures in the dynamical map with the help of a frequency analysis method. The tools and methods adopted here are very similar to those used in our investigation of Earth Trojans \citepads{2019A&A...622A..97Z}. Here is a brief introduction.

\subsection{Dynamical model and initial conditions}\label{subsec:modinit}
	
The outer solar system model consisting of the Sun (with the mass of inner planets added onto), four giant planets and massless fictitious UTs is adopted in our numerical simulations. Many studies for UTs and Neptune Trojans have verified the reliability of this model \citepads[see e.g.]{2010CeMDA.107...51D,2009MNRAS.398.1217Z,2011MNRAS.410.1849Z}. Since the phase spaces around $L_4$ and $L_5$ are dynamically symmetrical to each other \citepads[see e.g.]{2002Icar..160..271N,2003A&A...410..725M,2009MNRAS.398.1217Z,2019A&A...622A..97Z}, we just focus on $L_4$ in this paper.
	
We initialized the model with a planetary configuration at epoch of JD 245\,7400.5. The orbital elements of planets are taken from JPL HORIZONS system\footnote{\url{ssd.jpl.nasa.gov/horizons.cgi}} \citepads{1996DPS....28.2504G}. The fictitious UTs were initially put around $L_4$ sharing the same eccentricity $e$ {(0.04975)}, longitude of the ascending node $\Omega$ {(73.93$^\circ$)}, and mean anomaly $M$ with Uranus. {We set the initial argument of perihelion of all fictitious UTs as $\omega_0=\omega_7+60^\circ$}\footnote{The subscript `5' to `8' denotes planet Jupiter to Neptune respectively throughout this paper.}, {so that} the resonant angle of fictitious UTs, which is defined as
		\begin{equation}
     		\sigma=\lambda-\lambda_7\,,
   		\end{equation}
where $\lambda=\omega+\Omega+M$ is the mean longitude, is fixed at $60^\circ$ at the beginning of simulations. {It is worthy to note that the locations of $L_4$ and $L_5$ could be displaced from $\sigma=\pm 60^\circ$ owing to the presence of the eccentricity and inclination \citepads{2000CeMDA..76..131N,2007CeMDA..98..181K}, but in our case the deviation is not remarkable ($<5^\circ$).} To investigate the stability regions on the $(a_0,i_0)$ plane, a grid of initial conditions was considered with 271 initial values of semi-major axes and 131 values of inclinations equally spaced in the domain where $a_0$ ranges from 19.08\,AU to 19.35\,AU and $i_0$ ranges from $0^\circ$ to $65^\circ$ (35501 test particles in total).
	
We performed two primary sets of numerical simulations in our investigation, of which one integrated the whole system for $3.4\times10^7$\,yr while the other integrated for the age of the solar system (4.5\,Gyr). Adopting a Lie-series integrator \citepads{1984A&A...132..203H}, the short-time simulation was carried out to construct the dynamical map and locate the involved resonances. At the same time, a long-time simulation is expected to check the existence of primordial UTs with the help of the \textit{hybrid symplectic integrator} in the \textsc{mercury6} software package \citepads{1999MNRAS.304..793C}. {The orbits of planets remain stable in the long-time run, and specifically for Uranus, the maximum eccentricity during the simulation is smaller than 0.08.} The same integrator including an artificial force was also implemented to simulate the radial migration of planets. We note that the results from these two numerical tools are verified to be consistent with each other.
	 
\subsection{Spectral analysis and frequency analysis method}

It is necessary to apply a spectral analysis method in the short-time simulation to reduce the amount of output data and remove short-period terms which may cause some difficulties in the following frequency analysis.
	
Concretely, we made use of an on-line low-pass digital filter to smooth the output of the integration \citepads{1995A&A...303..945M,2002Icar..158..343M} and resampled the data with an interval of $\Delta=256$\,yr. For the whole integration time of $3.4\times10^7$\,yr, we could obtain $N=2^{17}$ $(=131\,072)$ lines of data for each orbit. Using the fast Fourier transform (FFT), we calculated the frequency spectra for each orbit and then recorded the spectral number (SN) which is defined as the number of peaks above a specific threshold in a spectrum. The SN is an appropriate indicator of orbital regularity carrying abundant information of the dynamical behaviour \citepads{2009MNRAS.398.1217Z,2011MNRAS.410.1849Z,2019A&A...622A..97Z}. Obviously, the smaller the SN, the more regular the orbit. In most cases, regularity could reflect stability, or even predict stability. The Nyquist frequency is $f_{\rm Nyq}=1/(2\Delta)=1.95\times10^{-3}\,2\pi\,{\rm yr}^{-1}$ while the spectral resolution is $f_{\rm res}=1/(N\Delta)=2.98\times10^{-8}\,2\pi\,{\rm yr}^{-1}$. Actually, all secular fundamental frequencies in the outer solar system except the nodal precession rate of Jupiter (see Sect.~\ref{subsec:dynspe}) are covered in the range of these two frequencies.
	
In order to locate the resonances responsible for the dynamical behaviour of UTs with a semi-analytical method (see Sect.~\ref{sec:fma}), we have to precisely determine the proper frequencies for each orbit with a frequency analysis method proposed by \citetads{1990Icar...88..266L}. For a brief introduction to the numerical algorithm, please refer to \citetads{2019A&A...622A..97Z}.

%-----------------------------------------------------------------

\section{Dynamical map}\label{sec:dymap}

Indicators such as the maximum Lyapunov characteristic exponent, diffusion rate in the frequency space, maximum eccentricity, libration amplitude and escape time are all well used in the study of orbital stability \citepads[see e.g.][]{2002Icar..160..271N,2003A&A...410..725M,2010CeMDA.107...51D}. In this paper, we adopt the SN of $\cos\sigma$ to construct a more detailed dynamical map. 
	
\subsection{Stability region}

\begin{figure*}
	\centering
	\includegraphics[width=9cm]{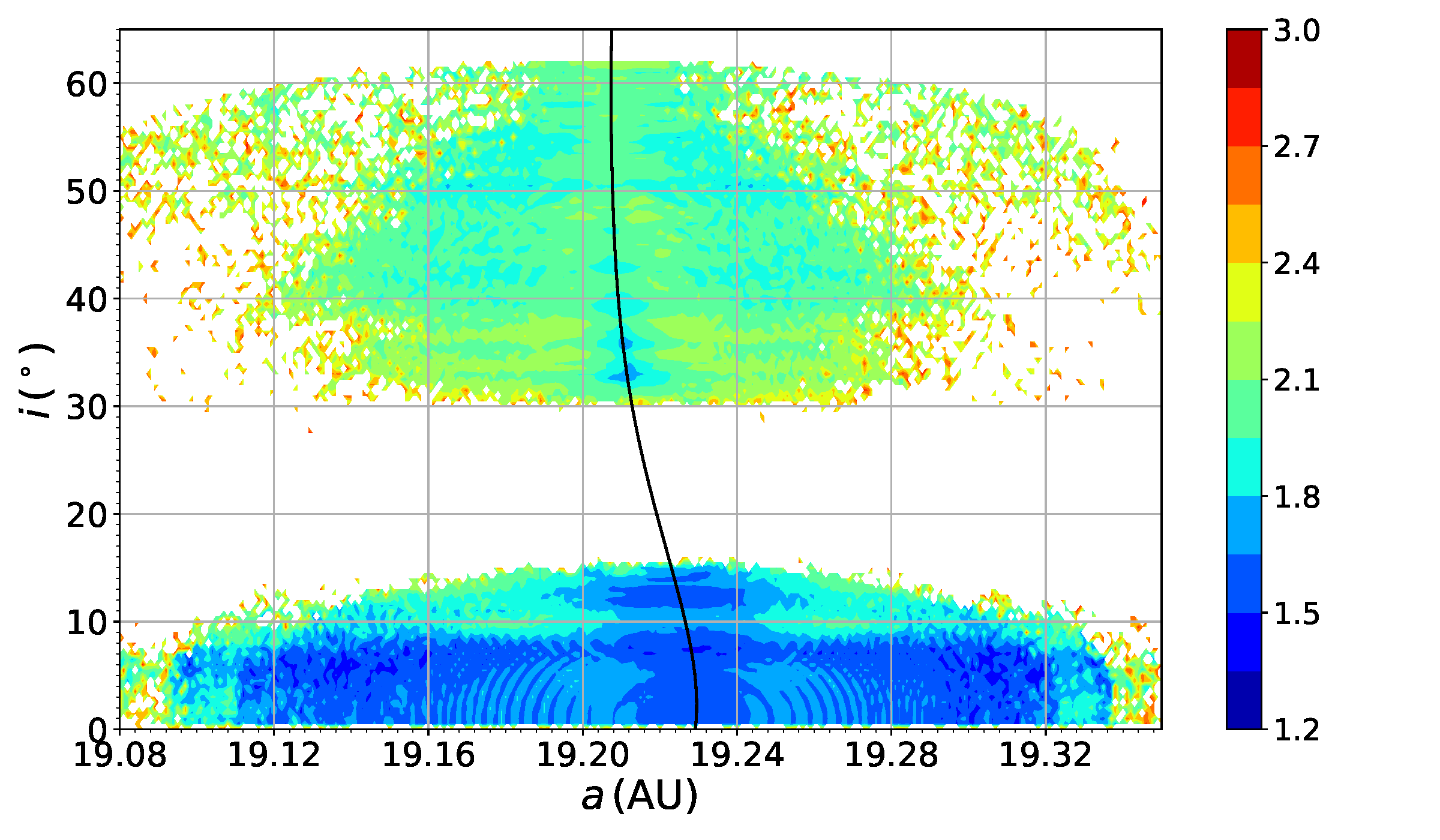}
	\includegraphics[width=9cm]{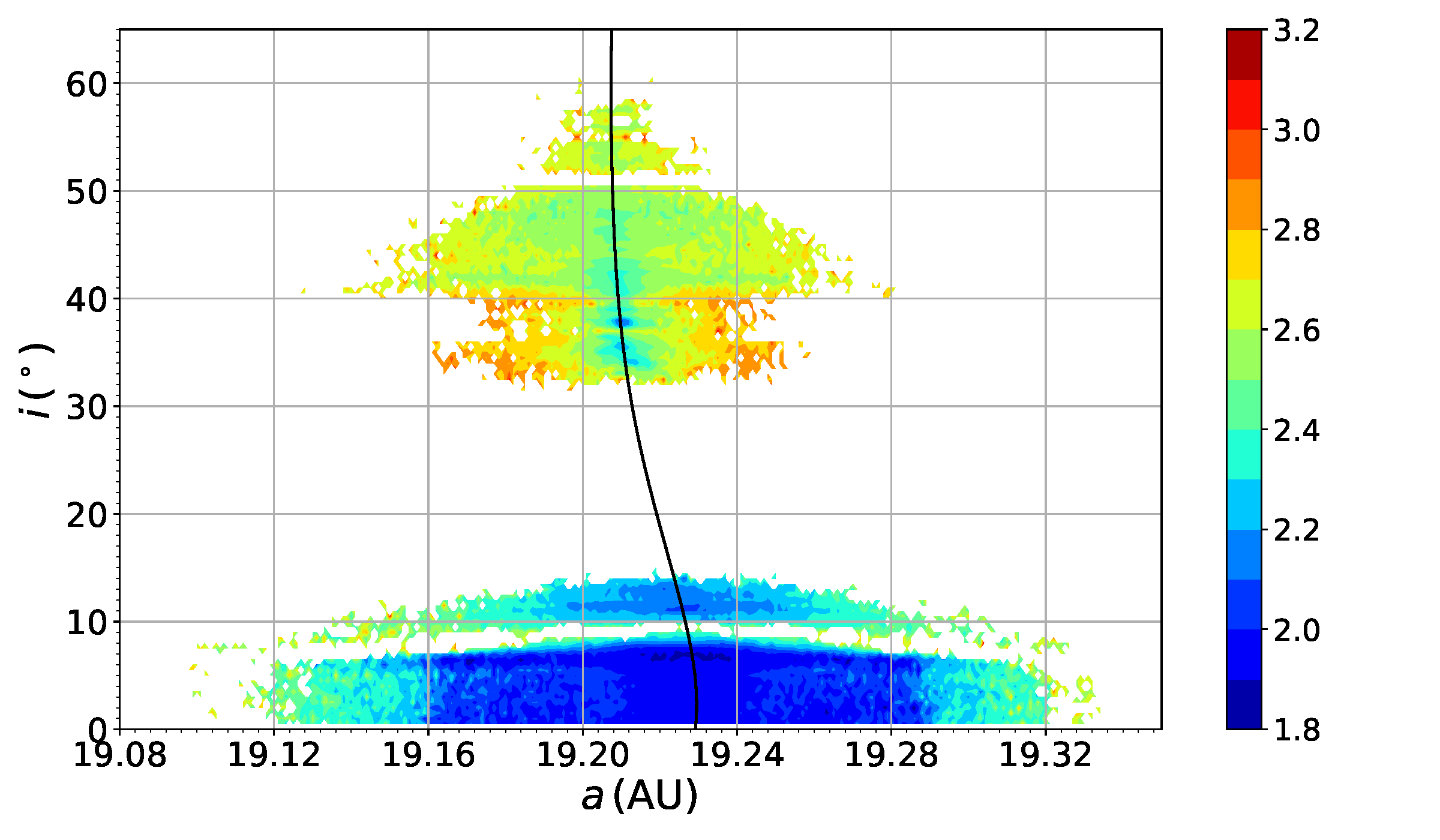}
	\caption{Dynamical maps around the $L_4$ point of Uranus on the $(a_0,i_0)$ plane. The colour indicating the regularity of orbits represents the base-10 logarithm of SN of $\cos\sigma$. Orbits escape from the 1:1 MMR with Uranus during the integration time are not included in these maps. The libration centres for different $i_0$ {are} depicted by the dark line. {The \textit{left panel} shows the result for $4.25\times10^6$\,yr while the \textit{right panel} is for the integration time ($3.4\times10^7$\,yr), which is eight times the former.}}
	\label{fig:sn}
\end{figure*}

We present the dynamical maps around $L_4$ of Uranus on the $(a_0,i_0)$ plane in Fig.~\ref{fig:sn}, where the colour represents the base-10 logarithm of SN. Results for two different integration time ($3.4\times10^7\,{\rm yr}$ and $4.25\times10^6\,{\rm yr}$) are separately shown. Red regions where orbits are of large values of SN indicate a poor regularity while blue regions are the opposite. Based on our test runs %and the separatrics around Uranus \citepads{2002CeMDA..82..323N}, 
we regard the orbits that dissatisfy $18.62\,{\rm AU} \le a \le 19.82\,{\rm AU}$ at any time during the integration as escaping from the co-orbital region of Uranus and they are excluded from the map. 

{In the explored region, stability islands are confined at two small regions on the $(a_0,i_0)$ plane (Fig.~\ref{fig:sn}), of which one is at low inclination ($0^\circ$--$14^\circ$) while the other is at high inclination ($32^\circ$--$59^\circ$).} Two nearly horizontal strips around $9^\circ$ and $51^\circ$ are found lying in these two stability regions respectively, dividing each of them into two regimes. For the low-inclined orbits, the stability region could extend to the range of $19.224\pm0.108\,{\rm AU}$ while it shrinks to $19.21\pm0.07\,{\rm AU}$ at high inclination. Obviously, the most stable orbits (in blue) whose SN is smaller than 200 reside mostly in the low-inclination regime. The subsequent long-time integration (see Sect.~\ref{sec:longt}) proves that these orbits have a great chance to survive the age of the solar system. It is worth noting that the center island of the high inclination window can also hold this kind of stable orbits, even though the high-inclined orbits are less stable than the low-inclined ones overall as they possess larger SN. %Naturally, the farther away from the centre in semi-major axis, the less stable the orbits. 

It can also be seen from the {right} panel in Fig.~\ref{fig:sn} that abundant horizontal fine structures appear in the stability regions, implying plenty of high-degree resonances hiding in the phase space. However, some high-degree resonances causing the patterns in the dynamical map may have a weak strength which can hardly affect the long-term stability of UTs. But still, they have a remarkable influence on the regularity of orbits and in turn, the fine structures could point us a way to locating the related resonances (see Sect.~\ref{sec:fma}). 
	
By the end of the whole integration time ($3.4\times10^7$\,yr), most orbits have escaped, leaving a large area of the $(a_0,i_0)$ plane blank. To explore these uninformative regions, we construct a dynamical map with the output of a shorter-time simulation ($4.25\times10^6$\,yr), where the integration time is only one-eighth of the previous one (the {left} panel in Fig.~\ref{fig:sn}). % Obviously, both stability regions expand, mainly in semi-major axis. 
Due to a much {shorter} evolution time, smaller SN is expected in this dynamical map as some long-term instabilities have not worked on the orbits yet. The instability gaps around $9^\circ$ and $51^\circ$ are absent, implying that the corresponding time scale of the resonance mechanism is longer than $4.25\times10^6$\,yr. On the contrary, the lack of orbits in moderate inclination suggests that the resonances residing between $15^\circ$ and $30^\circ$ are of shorter time scales. The formation of the cyclic structures apparently seen in the low-inclination regime is believed to be involved with short-period mechanisms such as some secondary resonances or three-body resonances (see Sect.~\ref{subsect:resweb}). However, these resonance mechanisms do not contribute to the long-term stability because the cyclic structures disappear after a period of time (see the {right} panel in Fig.~\ref{fig:sn}) and cause no escape of orbits there.

\subsection{Excitation of the eccentricity and inclination}\label{subsec:ext}

\begin{figure}
	\resizebox{0.9\hsize}{!}{\includegraphics{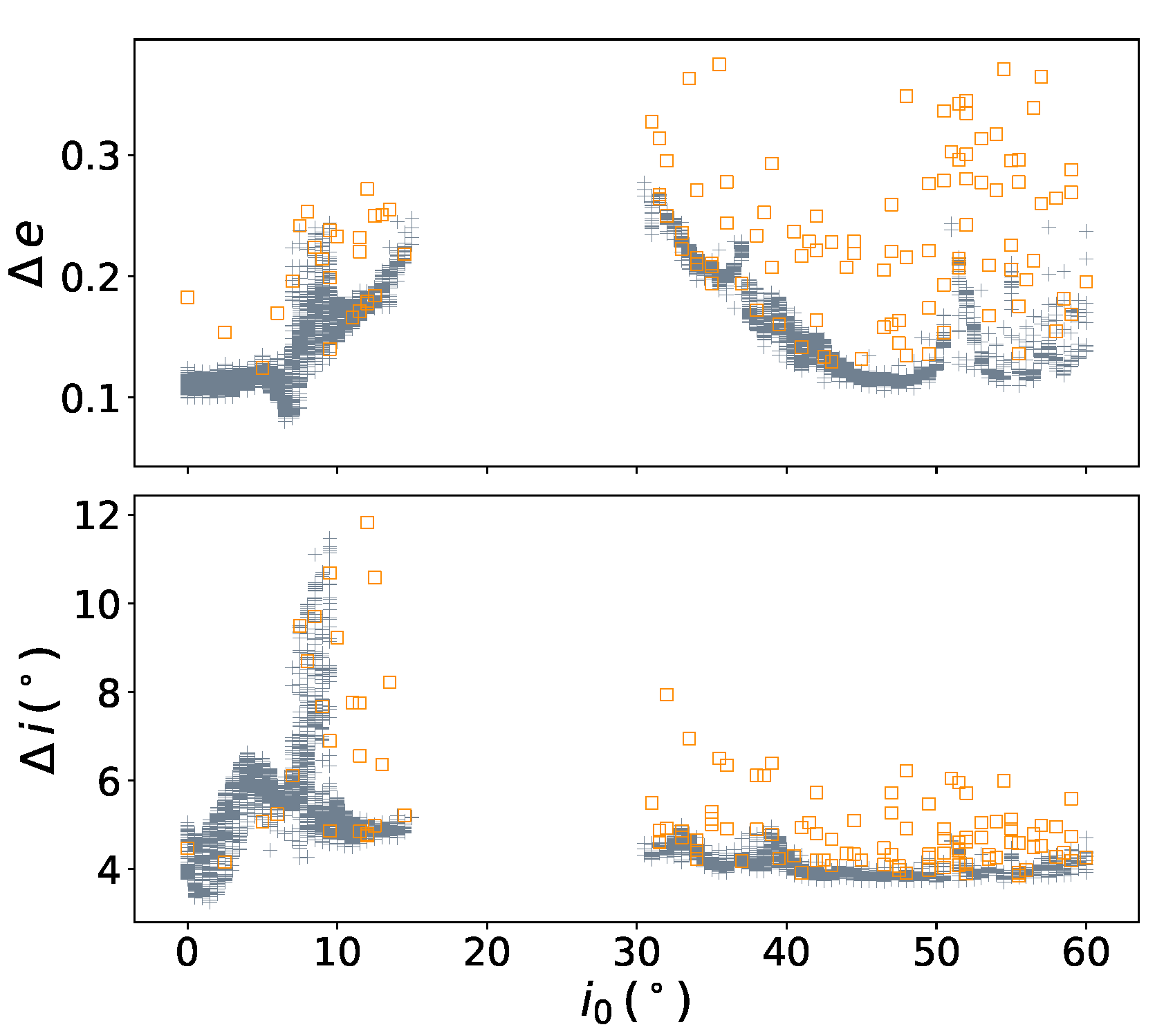}}
	\caption{Variations of the eccentricity and inclination for orbits with different initial inclinations surviving the integration time of $3.4\times10^7$ yr. {The initial eccentricity is 0.04975 for all orbits.} There might be many orbits with different initial semi-major axes for each initial inclination. The Orange squares indicate the orbits whose libration amplitude ($\Delta\sigma$) is larger than $300^\circ$ while black crosses indicate those which have never left the tadpole region around $L_4$ during the integration time ($\Delta\sigma<180^\circ$).}
	\label{fig:dedi}
\end{figure}

The variations of eccentricity and inclination can also provide some clues about related dynamical mechanisms. Since the variations of eccentricity and inclination are more dependent on initial inclination rather than initial semi-major axes, we show the amplitudes\footnote{{Throughout this paper,} by `amplitude’ we always mean the difference between the maximum and minimum values during the integration.} of eccentricity ($\Delta{e}$) and inclination ($\Delta{i}$) against initial inclinations ($i_0$) in Fig.~\ref{fig:dedi}. We find the maximum eccentricity of surviving orbits can hardly exceed 0.14, except those in the region from $i_0=9^\circ$ to $40^\circ$ and around $i_0=51^\circ$, where the eccentricity could reach 0.24 or even larger (the top panel in Fig.~\ref{fig:dedi}). The inclination of surviving orbits above $i_0=30^\circ$ mainly vary within $5^\circ$ while that of orbits around the instability gap at $i_0=9^\circ$ could be excited to $11.8^\circ$ (the bottom panel in Fig.~\ref{fig:dedi}). Besides, a peak of $\Delta{i}$ up to $6.5^\circ$ stands around $i_0=4^\circ$. We note there also exist some orbits diffusely distributed in Fig.~\ref{fig:dedi}, especially in the high-inclination regime. They are close to the boundary of the stability region and on the way of losing their stability. These orbits  with higher eccentricity and inclination are very likely to have left the tadpole region of $L_4$. In our simulations we find that {once} an object escapes from the tadpole cloud, it will obtain a quite large libration amplitude ($>300^\circ$, {mainly in horseshoe configuration) if it is still trapped in the 1:1 MMR} \citepads[see e.g. the phase portrait in][]{2007CeMDA..98..181K,2013CeMDA.117...17R}. We also note that no orbits are of libration amplitudes between $180^\circ$ and $300^\circ$ in our simulations.

As we know, the apsidal and nodal secular resonances are supposed to be responsible for the regional excitation of the eccentricity and inclination respectively according to the linear theory of secular perturbation \citepads[see e.g.][]{1999ssd..book.....M,2006ChJAA...6..588L}, so that Fig.~\ref{fig:dedi} could help us refine the search for the related resonances. The secular resonances inducing the instabilities between $14^\circ$ and $32^\circ$ are most probably involved with the perihelion precession, since Fig.~\ref{fig:dedi} infers that $\Delta e$ in this region is quite large. Furthermore, these apsidal secular resonances should have large width as orbits with a large range of inclinations are affected (see the top panel of Fig.~\ref{fig:dedi}). The orbits close enough to the secular resonances are able to be excited to $e>0.25$ and then lose their stability. Another apsidal secular resonance is found around $51^\circ$ with a much narrower width. It can drive the eccentricity up to $\sim0.21$. 
	
\begin{figure*}
	\resizebox{\hsize}{!}{\includegraphics{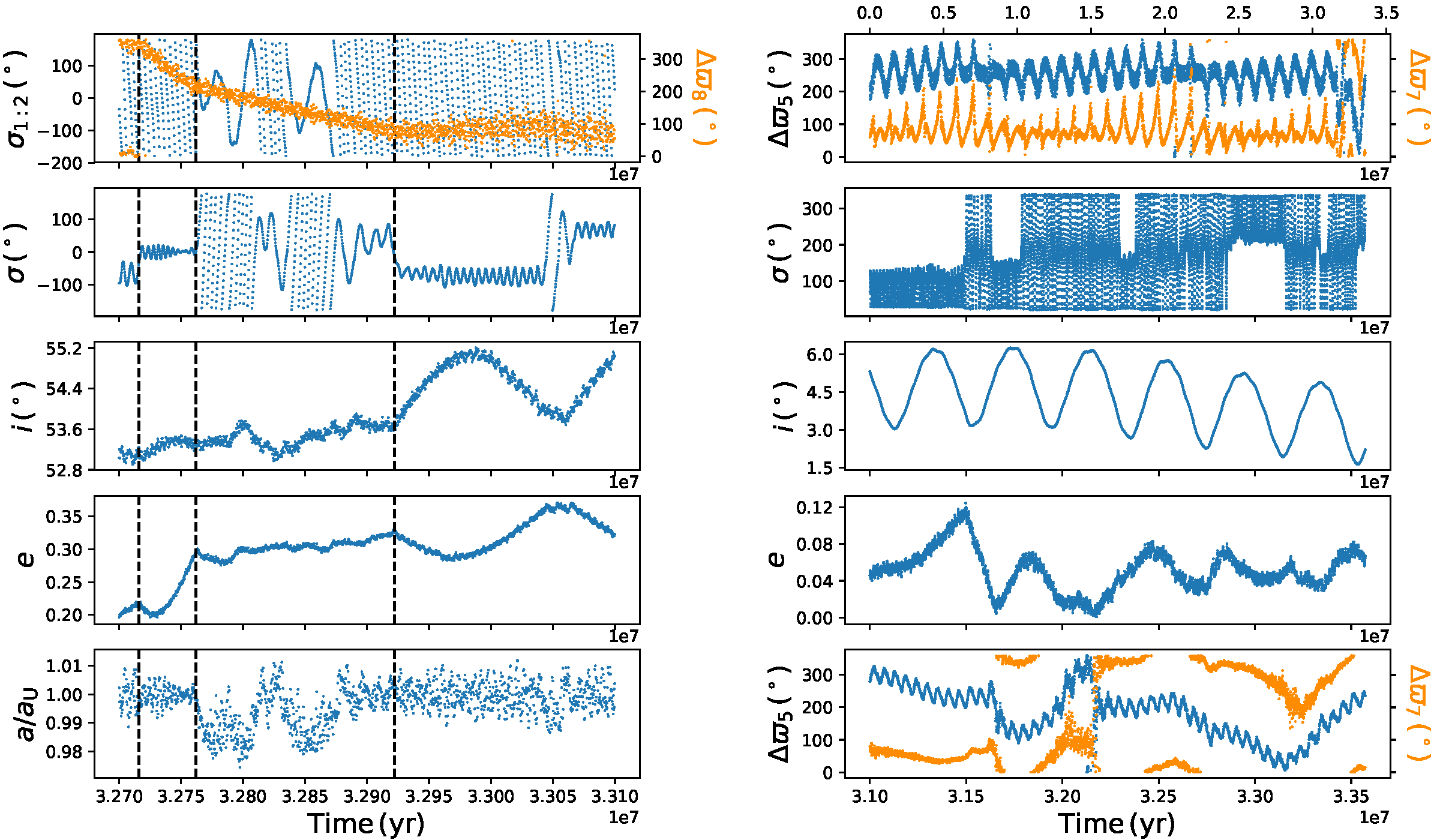}}
	\caption{Orbital evolution of two fictitious Trojans over some period at the end of the integration (except the \textit{top right} plot). The orbits in the \textit{left} and \textit{right panels} are initialized with $(a_0,i_0)=(19.225\,{\rm AU},56.5^\circ)$ and $(19.317\,{\rm AU},5^\circ)$ respectively. We illustrate the temporal evolution of the eccentricity $e$, inclination $i$ and resonant angle of the 1:1 MMR with Uranus $\sigma$ for both orbits. The {semi-major axis ratio of the UT and Uranus $a/a_{\rm U}$}, resonant angle of the 1:2 MMR with Neptune $\sigma_{1:2}$ and apsidal difference with Neptune $\Delta\varpi_8$ are also shown in the \textit{left panel} while the apsidal differences with Jupiter $\Delta\varpi_5$ and Uranus $\Delta\varpi_7$ over the last few million years (\textit{bottom right}) and the whole integration time (\textit{top right}) are shown respectively in the \textit{right panel}. The vertical dashed lines in the \textit{left panel} indicate the time $3.2716\times10^7\,{\rm yr}$, $3.2762\times10^7\,{\rm yr}$ and $3.2922\times10^7\,{\rm yr}$, when the UT enters and leaves the quasi-satellite orbit, and is captured by the Kozai mechanism. {The time is determined by checking if the resonant angle is librating around $0^\circ$ and if the argument of perihelion is librating around $\pm90^\circ$ with the eccentricity and inclination coupled to each other.}}
	\label{fig:obtexmp}
\end{figure*}

We monitor the orbital evolution and find {that} the most important secular resonances dominating the moderate-inclination regime are $\nu_5$ and $\nu_7$ while $\nu_8$ dominates the high-inclination regime (Fig.~\ref{fig:obtexmp}). In accordance with practice, we denote the apsidal and nodal secular resonances with the $j^{\rm th}$ planet as $\nu_j$ and $\nu_{1j}$ respectively. They will occur when $g=g_j$ and $s=s_j$, where $g$ and $s$ indicate the precession rate of perihelion and ascending node respectively. It is worth noting that the resonances around $i_0=7.5^\circ$ are able to influence both the eccentricity and inclination {(Fig.~\ref{fig:dedi})}, implying that they involve the precession of both the perihelion longitudes and ascending nodes. The inclination that can be excited up to $\sim 20^\circ$ experiences a secular variation with a period of tens of millions of years. Actually, a large portion of orbits over there has escaped and the surviving ones near the resonances will leave the 1:1 MMR with Uranus sooner or later. For lower inclinations, the orbits are strongly influenced by other nodal secular resonances.
	
The close encounter with Uranus when UTs are on large-amplitude horseshoe or quasi-satellite orbits could excite the eccentricity sporadically. The Kozai mechanism {inside the coorbital resonance \citepads{1999Icar..137..293N,2016MNRAS.460..966G}} is also likely to pump up the eccentricity of high-inclined orbits by trading the inclination, and in turn the inclination can also be driven up. These two mechanisms could also work together to excite orbits. There are few orbits surviving the whole integration time under the influence of these mechanisms, even if they survive, they are expected to leave the co-orbital region of Uranus soon. 
	
The orbits marked in orange square in Fig.~\ref{fig:dedi} undergoing huge variations in resonant angles ($\Delta\sigma>300^\circ$) are mainly excited by these two mechanisms in addition to secular resonances. We arbitrarily present one such example in the left panel in Fig.~\ref{fig:obtexmp}. The orbit is initialized with $(a_0,i_0)=(19.225\,{\rm AU},56.5^\circ)$ and is of a $\Delta{e}$ of 0.371. As we can see, at $3.2716\times10^7\,{\rm yr}$, the orbit transforms from the tadpole orbit around $L_5$ to quasi-satellite orbit as the resonant angle begins to librate around $0^\circ$. The eccentricity ends the periodic variation which is a result of the secular resonances (not shown in the figure) and undergoes a remarkable growth due to the close encounters with Uranus. From $3.2922\times10^7\,{\rm yr}$, the orbit has been trapped in the Kozai mechanism, leading to a further increase in eccentricity up to 0.371. In addition, the apsidal difference between Trojans and Neptune $\Delta\varpi_8$ is found in oscillation all the time, which means the $\nu_8$ secular resonance governs the variations of the eccentricity before the orbits are dominated by the Kozai mechanism. We note that the semi-major axis could librate around 18.94 AU ($a/a_{\rm U}\approx0.986$), where the 1:2 MMR with Neptune locates. We check the resonant angle $\sigma_{1:2}=\lambda-2\lambda_8+\varpi$ and find it oscillate around $0^\circ$ at that time. The resonance overlap between the 1:1 MMR with Uranus and 1:2 MMR with Neptune could induce chaos \citepads{1979PhR....52..263C}. 

Actually, some orbits with large libration amplitudes may maintain their relatively low eccentricity and inclination for a long time, %could not have been excited to high eccentricity or inclination, 
just like the orbit in the right panel in Fig.~\ref{fig:obtexmp}, whose initial semi-major axis and inclination is 19.317 AU and $5^\circ$ respectively. The amplitude of the eccentricity is only 0.124. Here we arbitrarily choose one orbit as the example since the evolution of this kind of orbits is very similar to each other. Reflected in the oscillation of $\Delta\varpi_5$ and $\Delta\varpi_7$ over the integration time, these orbits are supposed to be under the control of apsidal resonances with Jupiter ($\nu_5$) and Uranus ($\nu_7$) all the time. In fact, the rate of the apsidal precession of such UTs changes little in a wide range of initial orbital elements, implying that many orbits could be dominated by the same secular resonances (see Fig.~\ref{fig:dynspe}).	Although neither close encounter nor Kozai mechanism occurs to excite the eccentricity and inclination during the integration time, the large libration amplitude coming from the chaotically recurring transformations between tadpole and horseshoe orbits will most probably lead to close encounter with Uranus and depletion from the resonance region in future.

\subsection{Libration center}\label{subsec:libcen}

\begin{figure}
	\resizebox{\hsize}{!}{\includegraphics{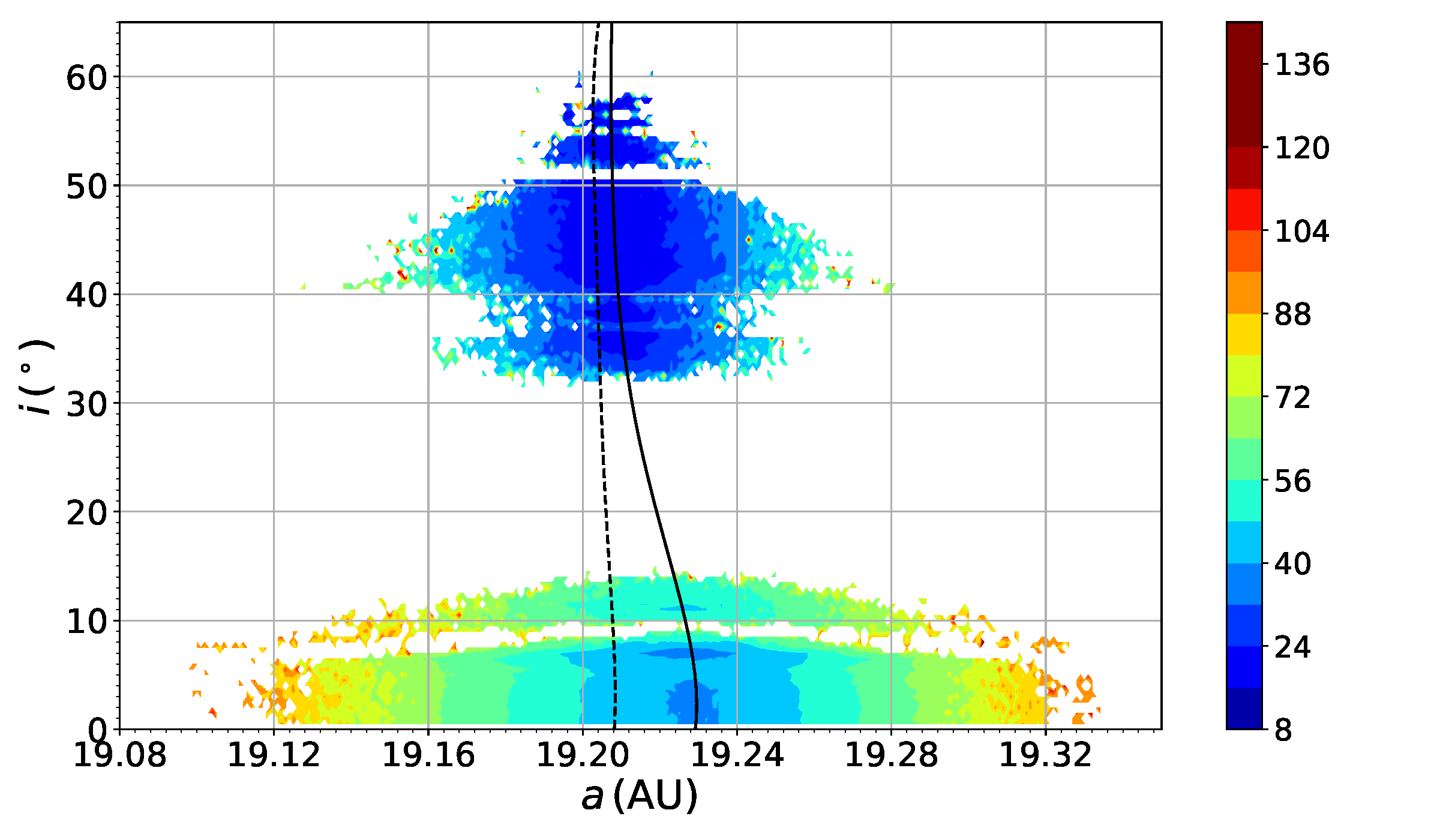}}
	\caption{Libration amplitudes of surviving orbits (in different colours) during the orbital integration of $3.4\times10^7$\,yr on the $(a_0, i_0)$ plane. The solid line implies the libration centers for different $i_0$. Dozens of orbits whose libration amplitude $\Delta\sigma$ is larger than $300^\circ$ are ignored for a better vision. Due to the scarcity of numbers, we indicate the orbits with libration amplitude $\Delta\sigma$ above $120^\circ$ with the same colour. {The dashed line is the same as the solid one, but for the integration where Neptune is initially placed further away from the 2:1 MMR with Uranus (see text).}}
	\label{fig:dsigma}
\end{figure}

In the planar circular restricted three-body model, test particles with small eccentricity are supposed to be on tadpole orbits if the radial separation from the secondary mass is less than $(8\mu/3)^{1/2}a$, where $\mu$ and $a$ {denote} the mass fraction and semi-major axis of the secondary mass \citepads{1999ssd..book.....M}. This criterion gives the Sun-Uranus system the maximum radial separation of tadpole orbits about 0.21\,AU. Although this range will shrink as the eccentricity increases, it is still wide enough to cover all stability regions shown in Fig.~\ref{fig:sn}, {given that UTs with long-term stability are limited to $e<0.1$} \citepads{2002Icar..160..271N}. It is consistent with the conclusion that horseshoe orbits of Uranus will lose their stability on short time spans because of the instabilities induced by the overlap of the 1:1 MMR with high-degree first-order MMRs \citepads{2002Icar..160..271N}. We show in Fig.~\ref{fig:dsigma} the libration amplitudes of surviving orbits on the $(a_0,i_0)$ plane. Most of the orbits in the stability region remain in the tadpole cloud around $L_4$ all through the integration time. But still, there exist some temporary horseshoe orbits with large libration amplitudes. They keep switching between different co-orbital states (see e.g. orbits in Fig.~\ref{fig:obtexmp}) and that is why there lack orbits whose $\Delta\sigma$ are between $180^\circ$ and $300^\circ$.

It is unexpected that the libration center shifts significantly with inclinations (see Fig.~\ref{fig:dsigma}) since this feature is absent for Earth Trojans and Neptune Trojans according to our previous investigations \citepads{2009MNRAS.398.1217Z,2011MNRAS.410.1849Z,2019A&A...622A..97Z}. We note that the libration center here is defined as the initial semi-major axis of the orbit whose libration amplitude is the smallest among the orbits with the same initial inclination. To depict the displacement, we numerically fit the libration center $a_c$ as a function of the initial inclination $i_0$ and show it in Fig.~\ref{fig:dsigma} as well as in the dynamical maps (Fig.~\ref{fig:sn}). Apparently, as the inclination increases, the libration center shifts towards smaller semi-major axis. This displacement is speculated to be involved with the quasi 2:1 MMR between Uranus and Neptune {(in current orbital configuration the orbital period ratio is 1.962)}. To confirm it, we initially placed Neptune at different distances from the exact location of the 2:1 MMR with Uranus. The simulations indicate that the displacement would gradually disappear when Neptune gets further away from the 2:1 MMR with Uranus {(see e.g. the dashed line in Fig.~\ref{fig:dsigma} for the case with period ratio of 1.945)}. In cases where Neptune is close enough to the 2:1 MMR, the overlap between the 1:1 MMR with Uranus and 1:2 MMR with Neptune could induce great chaos, destabilizing the UTs in a short time.
	
As we can also see from Fig.~\ref{fig:dsigma}, UTs with the smallest libration amplitudes are found in the high-inclination regime, where the SN is a bit larger than low-inclined UTs. It proves that the most stable UTs are of relatively larger libration amplitudes. The lack of low libration UTs in the low-inclination regime is due to the {cumulative} effect of indirect perturbations from other planets \citepads{2003A&A...410..725M}. The libration amplitude experiences a slight growth around $i_0=12^\circ$ before decreasing. Hence the stability island there by and large hosts the tadpole orbits of the largest libration amplitudes ($\sim52^\circ$ at the libration center). In fact, the secondary resonances (see Sect.~\ref{subsect:resweb}) could pump up the libration amplitude \citepads{1990Icar...85..394T,1990Icar...85..444M,2004Icar..167..347K}. These resonances, alongside with other important resonant mechanisms, will be figured out and located in next section with the frequency analysis method. %Thus it is necessary to find out the related resonances and locate them with the frequency analysis method.

%-----------------------------------------------------------------

\section{Frequency Analysis}\label{sec:fma}

\subsection{Dynamical spectrum}\label{subsec:dynspe}

\begin{figure}
  		\resizebox{0.9\hsize}{!}{\includegraphics{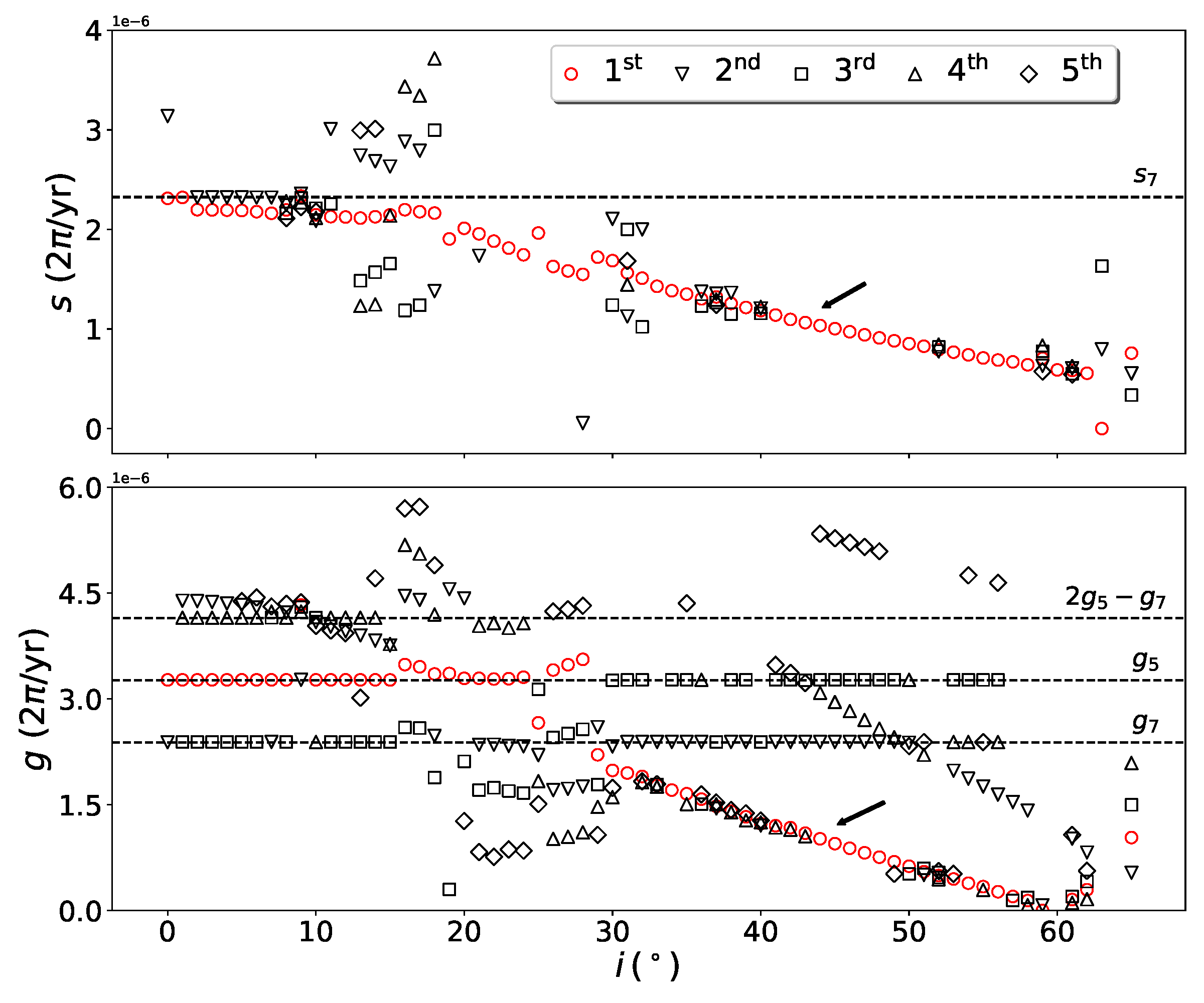}}
  		\resizebox{0.9\hsize}{!}{\includegraphics{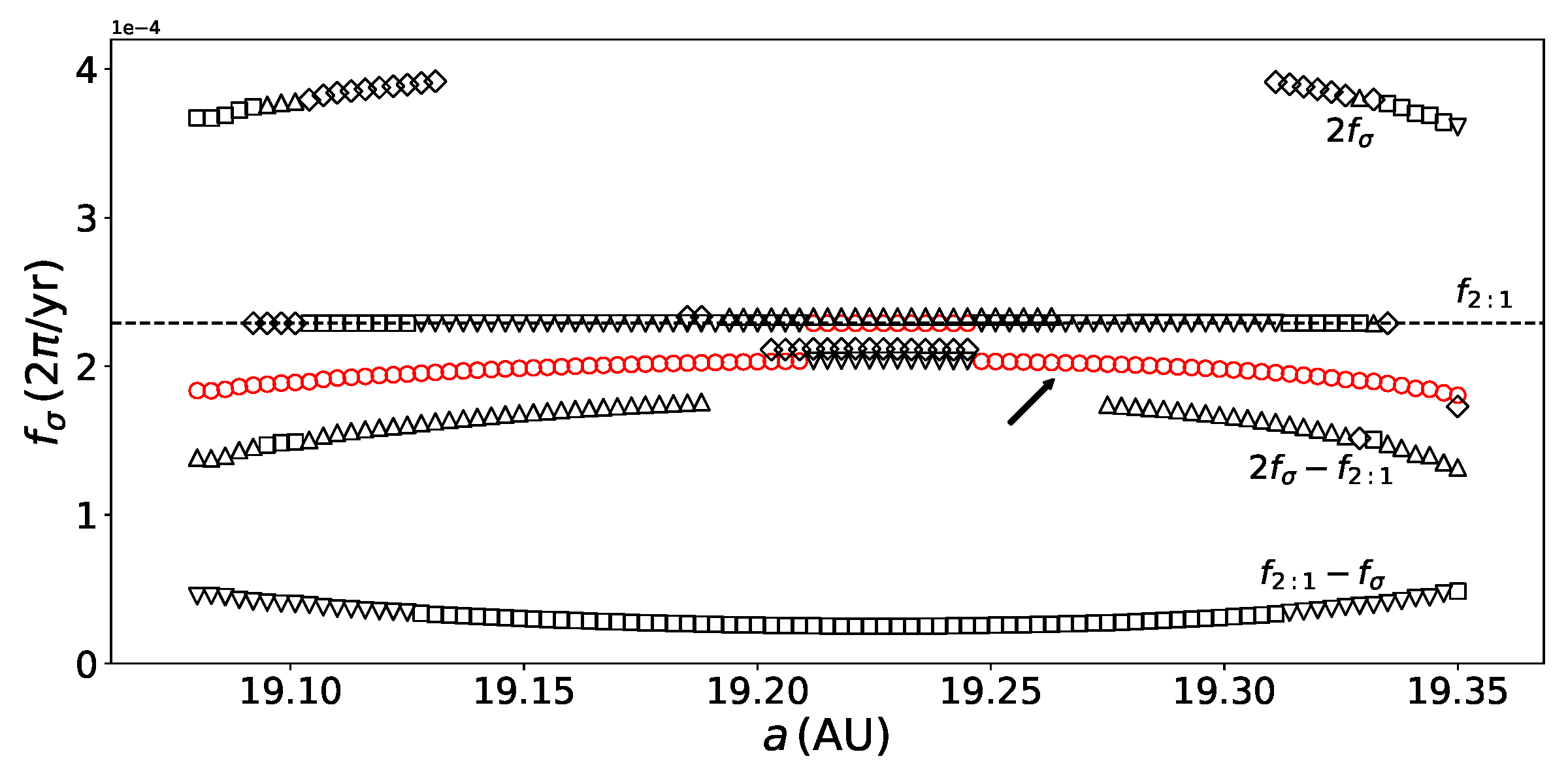}}
    		\caption{Dynamical spectra of $i\cos\Omega$ (\textit{top panel}), $e\cos\varpi$ (\textit{middle panel}) and $\cos\sigma$ (\textit{bottom panel}). The five most significant frequencies of the highest peaks in the power spectra, which are denoted by red open circles, black open inverted triangles, squares, triangles, and diamonds respectively, are plotted against their initial inclinations (\textit{top two panels}) or semi-major axes (\textit{bottom panel}). The proper frequencies {which vary continuously with the initial semi-major axis and inclination} are indicated by arrows. The initial semi-major axis for orbits in the \textit{top two panels} is 19.22\,AU while the initial inclination for orbits in the \textit{bottom panel} is $4^\circ$. The dashed lines indicate the forced frequencies which are identified to be involved with the frequency of the quasi 1:2 MMR between Uranus and Neptune $f_{2:1}$ or the fundamental frequencies in the outer solar system. In the \textit{bottom panel} the frequencies $2f$, $2f-f_{2:1}$ and $f_{2:1}-f$ are also labelled. We note that the precession rates of the ascending node for orbits in the \textit{top panel} are all negative as well as $s_7$. Hence we just plot the absolute values of these frequencies.}   		
    		\label{fig:dynspe}
\end{figure}

The frequency spectrum contains a lot of valuable information that can reflect the dynamical properties of orbits. Many significant components appear in the frequency spectrum of orbital variables, including the proper frequencies, forced frequencies, their harmonics and combinations. These frequencies can be computed precisely utilizing the method proposed by \citetads{1990Icar...88..266L}. We then picked out the proper frequencies via dynamical spectra, where several of the most important frequencies for each orbit are plotted against the initial semi-major axes or inclinations. Fig.~\ref{fig:dynspe} presents three arbitrary examples of dynamical spectra to illustrate the determination of the proper frequencies $f_\sigma$, $g$ and $s$ for the orbital variables $\cos\sigma$, $e\cos\varpi$ and $i\cos\Omega$ respectively. Apparently, the proper frequencies vary with the initial orbital elements while the forced frequencies remain unchanged. Their harmonics and combinations are also present.

\begin{table}[htbp]
		\centering
	\caption{Fundamental secular frequencies in the outer solar system. The data is computed from our simulation and agrees well with the results from \citetads{1989A&A...210..313N}.}
	\begin{tabular}{crrcrr}
	\hline\hline
  	& Period~~~ & Freq.~ &  & Period~~~ & Freq.~ \\
  	\hline
	$g_5$ & 306,433.17 & 32.63 & $s_5$ & $-129,548,180.$ & $-0.08$ \\
	$g_6$ & 46,831.03 & 213.53 & $s_6$ & $-49,128.01$ & $-203.55$ \\
	$g_7$ & 419,430.40 & 23.84 & $s_7$ & $-430,185.03$ & $-23.25$ \\
	$g_8$ & 1,917,396.1 & 5.22 & $s_8$ & $-1,864,135.11$ & $-5.36$ \\
  	\hline
	\end{tabular}
	\tablefoot{The periods are given in years and the frequencies are given in $10^{-7}\,2\pi\,{\rm yr}^{-1}$.}
	\label{tab:ffsolsys}
\end{table}

The precession rates of the perihelion ($g$) and ascending node ($s$) of Trojans are obscured by the forced frequencies which are identified to be associated with the fundamental secular frequencies in the outer solar system (Table~\ref{tab:ffsolsys}). The $s_7$ is found in the frequency spectra of $i\cos\Omega$ for low-inclined orbits while $2g_5-g_7$, $g_5$ and $g_7$ are present in the frequency spectra of $e\cos\varpi$. We note that the precession rate may be negative, just like the nodal precession rate of the orbits shown in Fig.~\ref{fig:dynspe} alongside the giant planets (see Table~\ref{tab:ffsolsys}). In the dynamical spectrum of $\cos\sigma$, the forced frequency is certified as the frequency of the quasi 1:2 MMR between Uranus and Neptune, which is denoted by $f_{2:1}=2.3238\times10^{-4}\,2\pi\,{\rm yr}^{-1}$ and corresponds to a period of 4303\,yr.

In the case of these three examples, the proper frequencies (indicated by arrows) often {have} the largest amplitudes (red open circles). However, this is not always the case and it is not necessary. The intersections of the proper and forced frequency are the exact locations where the secular resonances or secondary resonances take place {\citepads{1994IAUS..160..159M,2003MNRAS.345.1091M,2009MNRAS.398.1217Z,2011MNRAS.410.1849Z,2019A&A...622A..97Z}}. The corresponding proper frequencies nearby are affected and mired in a certain degree of confusion. For the same reason, the proper frequencies in the instability region are difficult to be identified. As we can see from Fig.~\ref{fig:dynspe}, the apsidal secular resonances with Jupiter and Uranus occupy the moderate-inclination regime. The nodal secular resonance with Uranus $\nu_{17}$ and apsidal secular resonance $g=2g_5-g_7$ could dominate the motion of low-inclined orbits. As we mentioned before in Sect.~\ref{subsec:libcen}, the quasi 1:2 MMR between Uranus and Neptune could have some influence on the dynamical properties of UTs.
	
\subsection{Resonance web}\label{subsect:resweb}

\citetads{2003A&A...410..725M} obtained the expression for the proper frequencies of UTs as a function of the proper orbital elements (eccentricity, inclination and libration amplitude). However, the dynamical maps in this paper are constructed in the osculating orbital elements space, which offers every convenience for the application of the maps by direct comparison with the observational data. Hence, instead of the proper elements, we numerically fit $f_{\sigma}$, $g$ and $s$ as a function of $a_0$ and $i_0$ (initial osculating elements) so that we can locate the resonances on the $(a_0,i_0)$ plane. % as soon as we pick out the proper frequencies of UTs. 
The quintic polynomial
	\begin{equation}
		f(x,y)=\sum_{m=0}^{5}\sum_{n=0}^{5-m}\,p_{mn}x^my^n
	\end{equation}
is adopted to fit the frequencies $f_{\sigma}$, $g$ and $s$, where $p_{mn}$ are the undetermined coefficients. The $x$ and $y$ represent the normalized $(a_0-a_c)$ and $\sin(i_0)$, where $a_c(i_0)$ is the libration center (see Sect.~\ref{subsec:libcen}). We note that the proper frequencies in chaotic areas are difficult to be determined so that the fitting formula may have some deviations in the margin area because the extrapolation may be relatively rough there. {A rough comparison shows that the values of proper frequencies obtained by \citetads{2003A&A...410..725M} and us are in similar ranges, especially in stability regions.}%due to a complete extrapolation.
	
For a more intuitive and comprehensive understanding of the dynamics in the phase space, we will display the resonances related to the orbital stability on the $(a_0,i_0)$ plane. The secular resonances can be located by solving the equation
	\begin{equation}
		pg+qs+\sum_{j=5}^{8}(p_jg_j+q_js_j)=0\,,	
	\end{equation}
where $p$, $q$, $p_j$ and $q_j$ are integers to be solved. The d’Alembert rule $p+q+\sum_{j=5}^{8}(p_j+q_j)=0$ has to be satisfied and $(q+\sum_{j=5}^8q_j)$ must be even. The value of $|p|+|q|+\sum_{j=5}^{8}\left(|p_j|+|q_j|\right)$ is the degree of the secular resonance.
	
\begin{figure}
	\centering
  		\resizebox{\hsize}{!}{\includegraphics{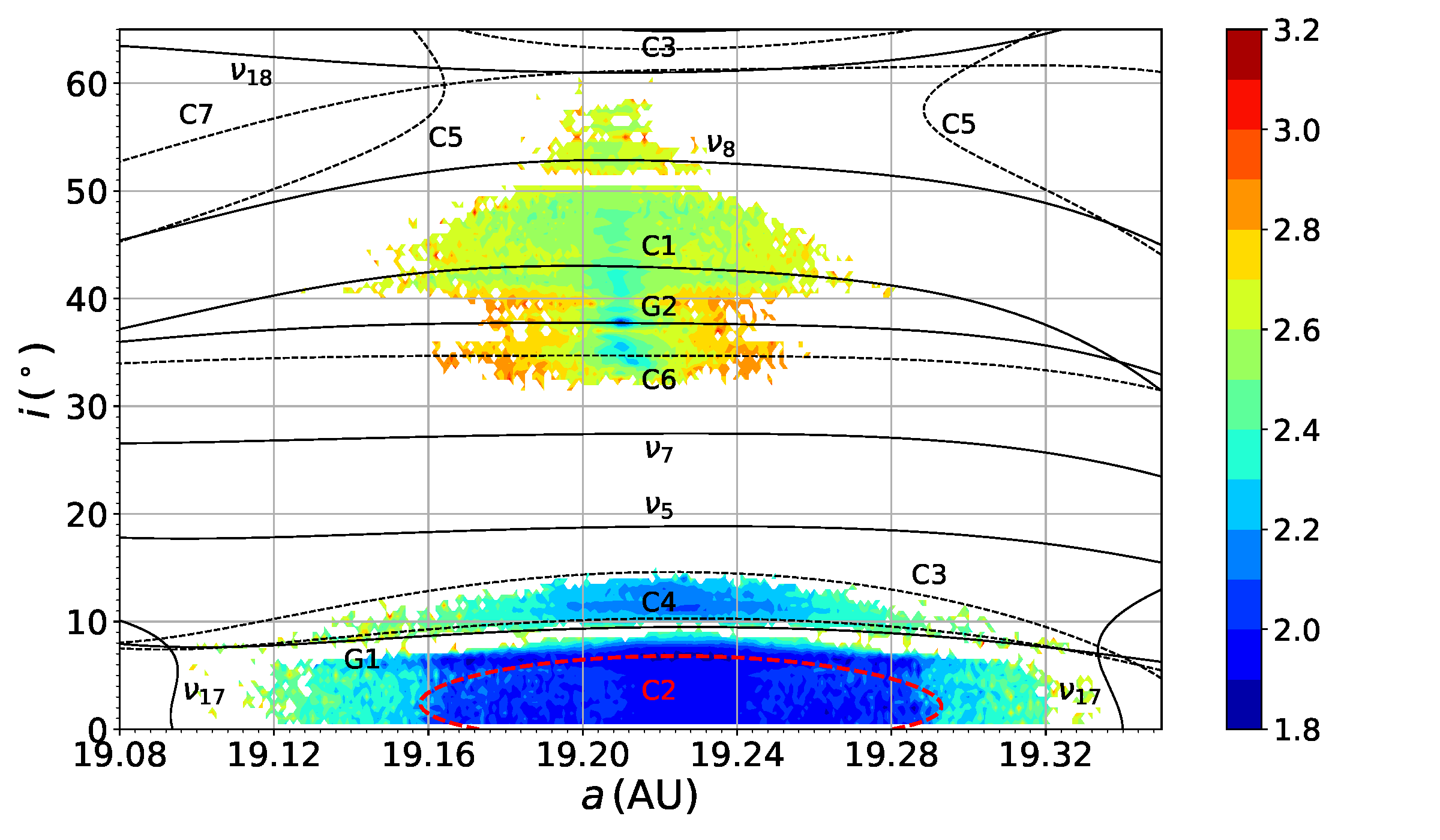}}
    		\caption{Locations of secular resonances for Trojans around $L_4$ on the $(a_0,i_0)$ plane. The resonances are labelled (see text for the meaning) along the curves. The solid lines indicate the linear and fourth-degree secular resonances while the dashed lines indicate the higher-order ones. The C2 resonance is in red for a better vision.}
    		\label{fig:gsres}
	\end{figure}

After a comprehensive search, we show the significant secular resonances in Fig.~\ref{fig:gsres}. Linear secular resonances are the most important mechanisms shaping the stability region. The $\nu_5$ and $\nu_7$ are depicted in the moderate-inclination regime, clearing the orbits from $14^\circ$ to $32^\circ$. As mentioned in Sect.~\ref{subsec:ext}, these two secular resonances have large widths on the $(a_0,i_0)$ plane. UTs close to them can be excited to high-eccentric orbits. The overlap between them can also induce chaos. $\nu_8$ could give rise to the instability strip around $51^\circ$. Although the exact locations of $\nu_{17}$ and $\nu_{18}$ are far from the stability region, they are still involved with the low-inclined and high-inclined orbits over there, controlling the variation of the inclination.
	
Following the classification of \citetads{2019A&A...622A..97Z}, we name the higher-degree secular resonances only involving the apsidal precession ($g$) ``G'' type and those only involving nodal precession ($s$) ``S'' type. The rest is called ``C'' type in the name of ``combined''. The meaning of these resonances shown in Fig.~\ref{fig:gsres} is explained as follows:
	\begin{equation}
  	\label{eqn:gsres}
  	\begin{aligned}%{array}{llll}
    {\rm G1}:  &  g-2g_5+g_7=0,  &
    {\rm G2}:  &  2g-g_7-g_8=0, \\
    {\rm C1}:  &  g+s-g_8-s_8=0, &
    {\rm C2}:  &  g+2s-s_5-g_8-s_8=0, \\
    {\rm C3}:  &  g+2s-g_8-2s_8=0, &
    {\rm C4}:  &  2g+s-2g_5-s_8=0, \\
    {\rm C5}:  &  2g+s-3s_5=0, &
    {\rm C6}:  &  2g-s+s_5-2g_7=0, \\
    {\rm C7}:  &  2g-3s+s_7=0.
  	\end{aligned}
	\end{equation}	
The G-type resonance $g-2g_5+g_7=0$ (G1) is supposed to be responsible for the instability gap around $i_0=9^\circ$. At the same time, C4 reside in the similar location. \citetads{2003A&A...410..725M} found that the forced frequency $2g_7-g_5$ could be {commensurate} with the apsidal precession rate $g$ for high-inclined orbits. In fact, G2 and $g-2g_7+g_5=0$ are almost located in the same position because $(g_7+g_8)/2\simeq2g_7-g_5$. Therefore, it should be noted that many resonances could share the same location on the $(a_0,i_0)$ plane and they can act together on the orbits. We classify these resonances into the same resonance family and just show some representative resonances in Fig.~\ref{fig:gsres}. Sometimes, the strength of the resonances belonging to the same family differs greatly. Then the motion is dominated by the strongest one. In other cases, these resonances are of similar strength, which may {lead} to instability as a result of the resonance overlap. The horizontal fine structures are related to the high-degree secular resonances such as C1, G2 and C6. Secular resonances in C2 family are believed to control the variation of the inclination for low-inclined orbits, causing the structure around $i_0=7.5^\circ$ in the bottom panel of Fig.~\ref{fig:dedi}. In addition, C3-like resonances could shape the outline of the stability region in the low-inclination regime.
	
\begin{figure}
  		\resizebox{\hsize}{!}{\includegraphics{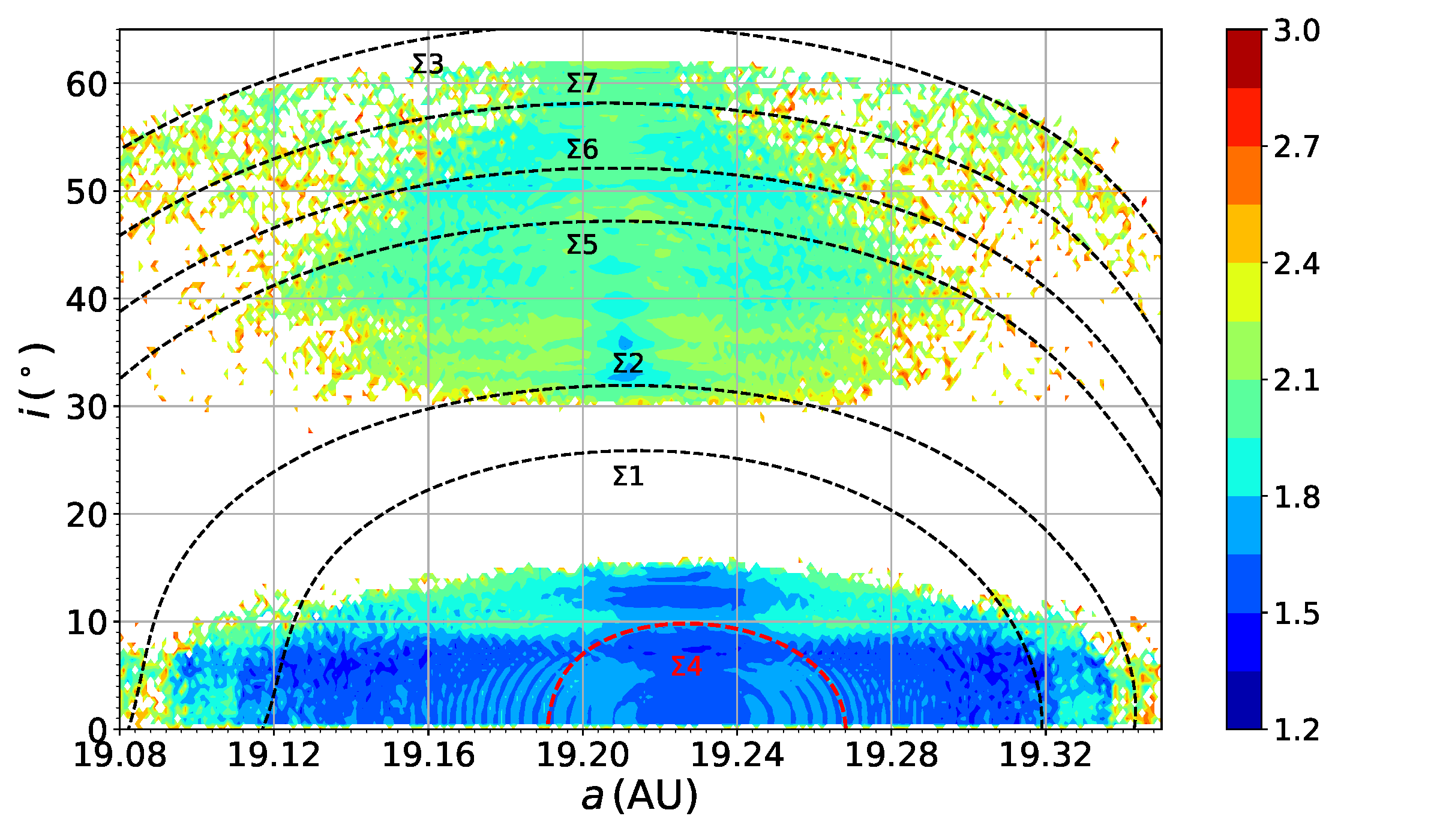}}
  		\caption{Same as Fig.~\ref{fig:gsres} but for {examples} of secondary resonances. We note that the dynamical map behind the resonances is the one for the integration of $4.25\times10^6$\,yr. The $\Sigma4$ resonance is in red for a better vision.}
  		\label{fig:fres}
\end{figure}
	
Following the terminology used by \citetads{1990Icar...85..444M}, the secondary resonances for UTs are those resonances that arise when the libration frequency of the resonant angle for the 1:1 MMR ($f_\sigma$) is commensurate with the circulation frequency of the resonant angle for the quasi 1:2 MMR between Uranus and Neptune ($f_{2:1}$). They can be expressed as
	\begin{equation}
	\label{equ:sec}
	hf_\sigma+kf_{2:1}\sim0, 
	\end{equation}
where $h$ and $k$ are integers.
Three-body resonances involving UTs, Uranus and Neptune are defined by the relation
	\begin{equation}
	\label{equ:tbr}
	m_T\dot{\lambda}+m_U\dot{\lambda_U}+m_N\dot{\lambda_N}\sim0, 
	\end{equation}
where $m_T$, $m_U$ and $m_N$ are integers and $\lambda$, $\lambda_U$ and $\lambda_N$ are the mean longitudes of UTs, Uranus and Neptune respectively. We can find that for UTs, three-body resonances are almost equivalent to secondary resonances if we could rewrite the resonant angle of three-body resonances as $m\,(\lambda-\lambda_U)+n\,(\lambda_U-2\lambda_N)$, where $m$ and $n$ are integers. It is valid because both $f_\sigma$ and $f_{2:1}$ are small so that Equation~(\ref{equ:tbr}) can be satisfied.
		
We search secondary resonances by solving
	\begin{equation}
	\label{equ:sec2}
		hf_\sigma+kf_{2:1}+pg+qs+\sum_{j=5}^{8}(p_jg_j+q_js_j)=0\,.	
	\end{equation}
The d’Alembert rule requires $k+p+q+\sum_{j=5}^{8}(p_j+q_j)=0$ in this case. Note that Equation~(\ref{equ:sec2}) is consistent with Equation~(\ref{equ:sec}) because the terms involving the secular precessions are much smaller than those involving $f_\sigma$ and $f_{2:1}$. Representative secondary resonances shown in Fig.~\ref{fig:fres} are listed as follows:
	\begin{equation}
  	\label{eqn:fres}
  	\begin{aligned}%{array}{llll}
    {\rm \Sigma1}:  &  f_\sigma-f_{2:1}-g+2g_6=0, &
    {\rm \Sigma2}:  &  f_\sigma-f_{2:1}-2s+2g_6+g_8=0, \\
    {\rm \Sigma3}:  &  2f_\sigma-f_{2:1}-g_6+2s_6=0, &
    {\rm \Sigma4}:  &  2f_\sigma-2f_{2:1}-g+3g_6=0, \\
    {\rm \Sigma5}:  &  3f_\sigma-2f_{2:1}+2s_6=0, &
    {\rm \Sigma6}:  &  3f_\sigma-2f_{2:1}+s_6+s_7=0, \\
    {\rm \Sigma7}:  &  3f_\sigma-2f_{2:1}+2s_8=0.
  	\end{aligned}
	\end{equation}
{Since the shapes match well,} the cyclic structures in the low-inclination regime are {speculated} to be caused by those secondary resonances. {Note that we show in Fig.~\ref{fig:fres} only one corresponding resonance ($\Sigma4$) for the cyclic structures, which are caused by plenty of such high-order secondary resonances. Since they involve $f_{\sigma}$ and $f_{2:1}$ whose time scales are relatively short, the secondary resonances are expected to affect the motion in relatively short term. In fact, the cyclic structures in the low-inclination regime can be found only in the dynamical map of short integration time $(4.25\times 10^6$\,yr) but not in the long term integration (see Fig.~\ref{fig:sn}). }
	
The secondary resonances or three-body resonances can affect the stability of Trojans in the outer solar system to varying degrees \citepads{2003MNRAS.345.1091M,2003A&A...410..725M,2009MNRAS.398.1217Z,2011MNRAS.410.1849Z}. Studies of the uranian satellite system indicate that orbits captured in the secondary resonances may experience an amplification of the libration amplitude of the primary resonance \citepads{1990Icar...85..394T,1990Icar...85..444M,2004Icar..167..347K}, which in our case is the 1:1 MMR with Uranus. The amplification will eventually lead to the escape from the co-orbital region, implying the secondary resonances could destabilize the orbits. However, our investigation suggests that the secondary resonances associated with $f_{2:1}$ have no significant effect on the long-term stability of UTs in the current planetary configuration although they could pump up the libration amplitude mainly in the low-inclination regime (see Fig.~\ref{fig:dsigma}). This is different from Neptune because MMRs exterior to a planet (1:2 MMR of Uranus) are supposed to be wider than MMRs interior to it (2:1 of Neptune), just like the different influence of the quasi 2:5 MMR between Jupiter and Saturn (the Great Inequality) on Jupiter Trojans and Saturn Trojans \citepads{2002Icar..160..271N}. As a result, the quasi 1:2 MMR between Uranus and Neptune could modify the structure of the phase space {on short time spans} rather than affecting the long-term stability of UTs. {\citetads{2016CeMDA.126..519P} and \citetads{2018CeMDA.130...20P} developed the `basic Hamiltonian model' to analytically determine the centre and boundary of secondary resonances in the space of proper elements for co-orbital motion and demonstrated that the innermost boundary of the most conspicuous secondary resonance could delimit the effective stability region for Trojans.}

%-----------------------------------------------------------------

\section{Long-term Stability }\label{sec:longt}
	
\begin{figure}
	\resizebox{\hsize}{!}{\includegraphics{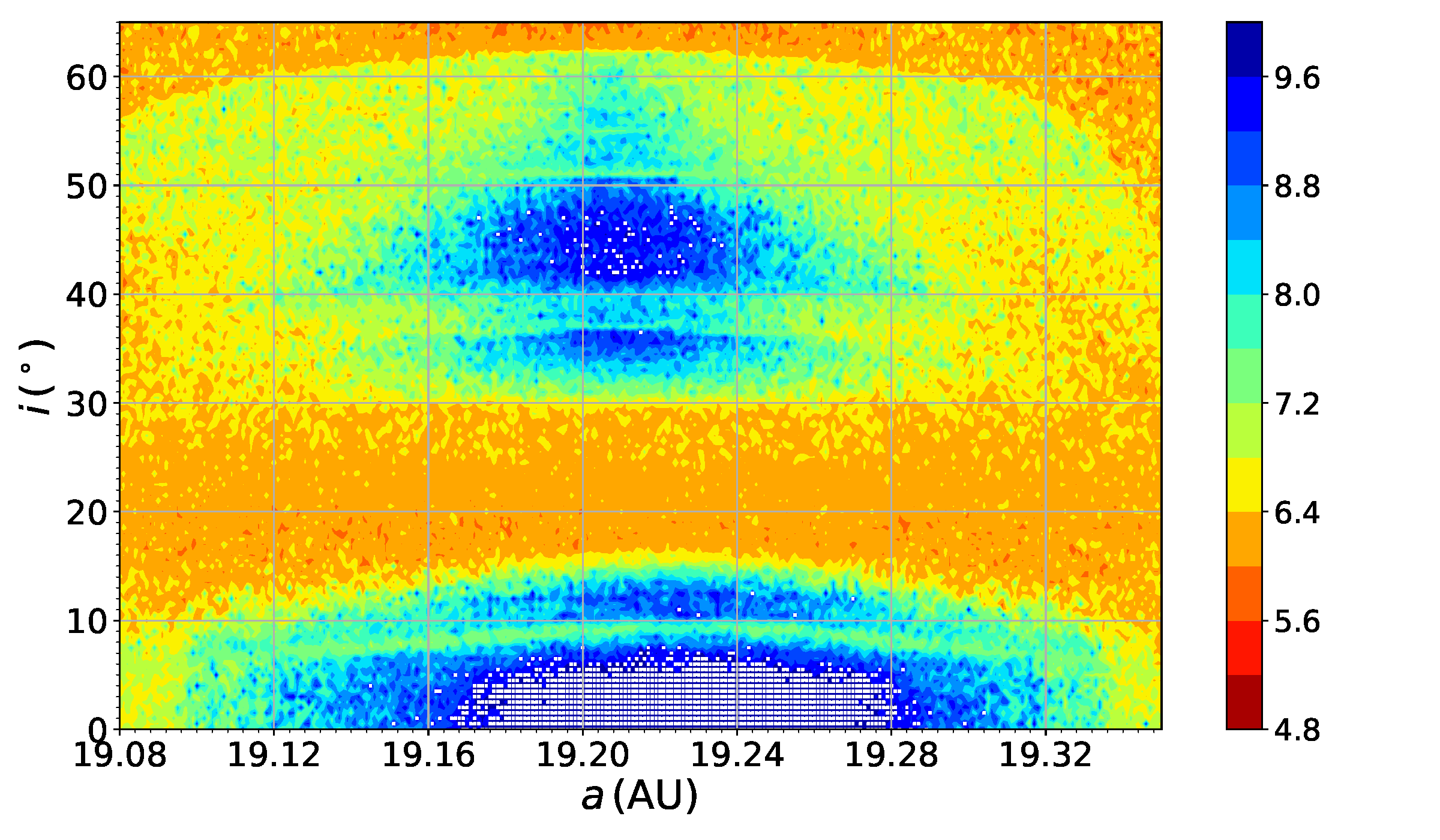}}
	\caption{Lifespans of fictitious UTs around the $L_4$ point on the $(a_0,i_0)$ plane. The colour indicates the {base-10} logarithm of the lifespans. The white points indicate the orbits surviving the age of the solar system (4.5\,Gyr).}
	\label{fig:lf}
\end{figure}

To assess the long-term stability of UTs, we integrated all the 35\,501 fictitious Trojans on the $(a_0,i_0)$ plane (see Sect.~\ref{subsec:modinit}) to the age of the solar system (4.5\,Gyr).  For each of them, the time when the orbit escapes from the co-orbital region of Uranus is recorded as its lifespan. Fig.~\ref{fig:lf} presents the lifespans of all orbits on the $(a_0,i_0)$ plane. The orbits of long lifespan form similar structures as the dynamical maps (Fig.~\ref{fig:sn}). {The cap structure around $60^\circ$ that is only obvious in the dynamical map for short-time integration (see the left panel of Fig.~\ref{fig:sn}) appears again. It} may be related to the C7-like secular resonance. The fine structures which could help us locate the resonances are absent in Fig.~\ref{fig:lf}. This is because the SN characterizes the regularity of the orbit and is not exactly equivalent to the lifespan in co-orbital region. Lifespan can intuitively describe the stability of orbits but SN can better reflect the characteristics of motion.

Among all fictitious UTs, about 3.81\% orbits survive the integration time. These stable orbits mostly (95.5\%) gather in the low-inclination regime around the libration center ($i_0<7.5^\circ,\,19.164\,{\rm AU}<a<19.288\,{\rm AU}$), which is consistent with the stability region where the orbits with the smallest SN reside. The rest (4.5\%) surviving orbits are in the high-inclination regime, from $42^\circ$ to $48^\circ$. {Some UTs with inclination up to $60^\circ$ could reside in the tadpole cloud around $L_4$ for $10^8$ yr. It is consistent with the analysis of periodic orbits around $L_4$ in \citetads{2016MNRAS.460..966G}.}  \citetads{2002Icar..160..271N} suggested that dynamical instabilities in the long-term evolution over 4\,Gy could cause a 99\% depletion of primordial UTs and Saturn Trojans while it is only 50\% for Neptune Trojans. The survival ratio (3.81\%) we obtain is larger than their result (1\%) because they only sampled $i_0=0^\circ$, $5^\circ$, $10^\circ$, $15^\circ$, $20^\circ$ and $25^\circ$ and the stability regions only exist for the first two conditions according to Fig.~\ref{fig:sn}.

	\begin{figure}
	\centering
  		\resizebox{\hsize}{!}{\includegraphics{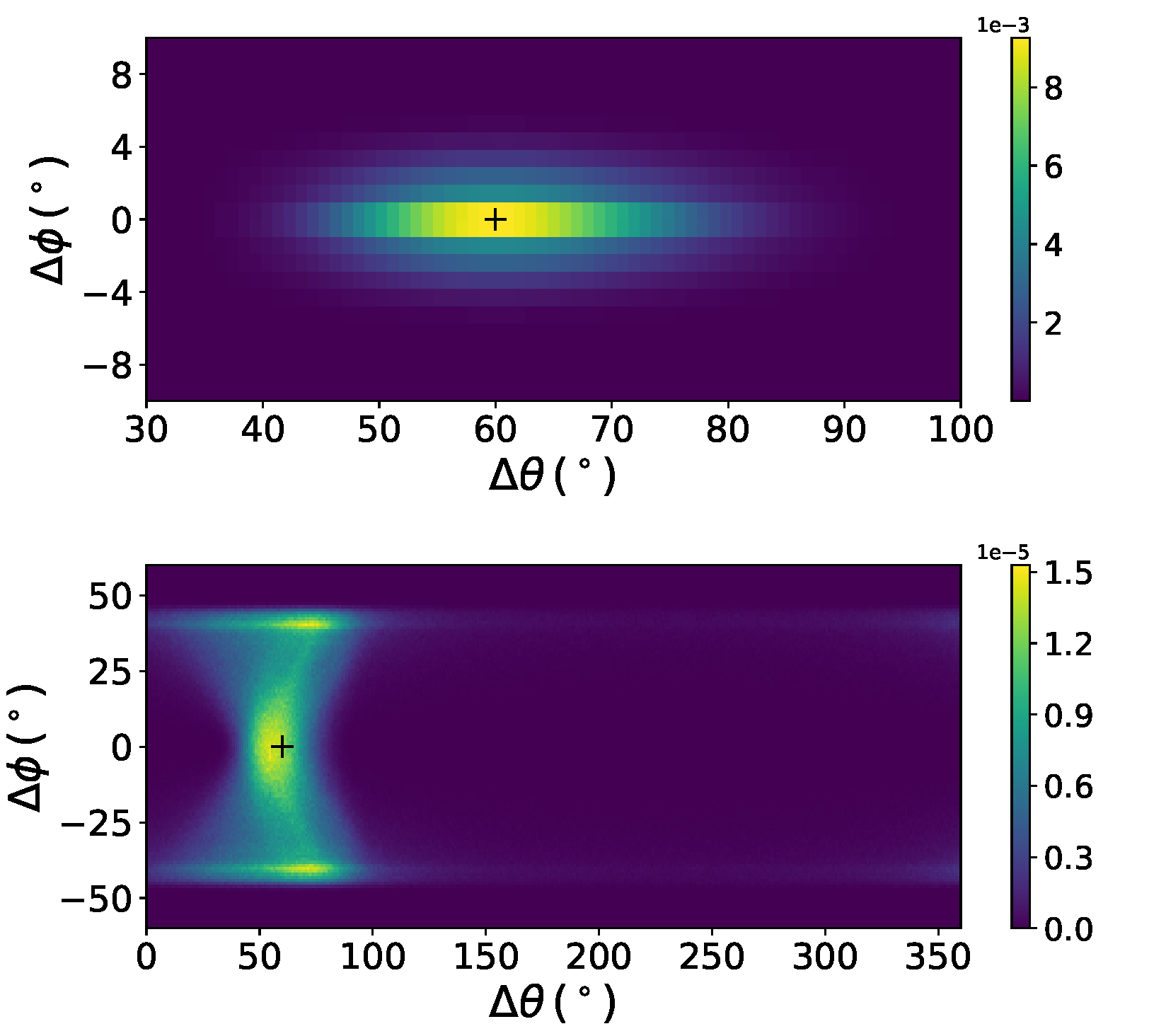}}
    		\caption{The normalized sky densities computed from the orbits surviving the age of the solar system. The \textit{top panel} is for {low-inclined} orbits while the \textit{bottom panel} is for {high-inclined} orbits. The colour represents the proportion of the residence time in each $1^\circ\times1^\circ$ segment of the sky with respect to the instantaneous orbital plane of Uranus. The black cross marks the position of the $L_4$ point.}
    		\label{fig:sky}
	\end{figure}
	
To guide the search for primordial UTs, we calculated the normalized sky densities making use of the orbital evolution over the age of the solar system. All surviving orbits are taken into account. The sky positions are referred to the instantaneous orbital plane of Uranus. Then we computed the relative longitudes ($\Delta\theta$) and latitude ($\Delta\phi$) for all orbits. The sky densities are obtained by calculating the time proportion that all orbits spend in each $1^\circ\times1^\circ$ segment of the sky. This is equivalent to the probability of UTs appearing in each segment at a certain time.
	
The result is shown in Fig.~\ref{fig:sky}. {Considering the high-inclined orbits ($i_0\in [42^\circ, 48^\circ]$) are much fewer than the low-inclined ones ($i_0<7.5^\circ$, see Fig.~\ref{fig:lf}), we presented their sky densities separately.} However, both of them are still normalized by the total residence time of all orbits. {Thus the sky densities resulted from them are very different from each other, as indicated by the colour bars with significantly different values in Fig.~\ref{fig:sky}.} Apparently, for low-inclined orbits, most of primordial UTs gather in the interval of $\pm5^\circ$ of the orbital plane of Uranus and they are most likely to be found on the coplanar orbits, especially where the $L_4$ point locates {$(\Delta\theta,\Delta\phi)=(60^\circ,0^\circ)$}. For high-inclined orbits, the largest densities occur within a region of $\sim$480 deg$^2$ around $(\Delta\theta,\Delta\phi)=(55^\circ,0^\circ)$. Besides, $(\Delta\theta, \Delta\phi) = (72^\circ,\pm40^\circ)$ is a good choice for observations dedicated to high-inclined primordial UTs. We note that the sky density in Fig.~\ref{fig:sky} is different from the one obtained by \citetads{2002Icar..160..271N} mainly because they have much fewer samples (only 14 orbits) with different distribution of orbital elements.
		
%-----------------------------------------------------------------

\section{Influence of Planetary Migration}\label{sec:pltmig}

\subsection{Pre-formed UTs}

{Although it may be related to the detection capacity, the} null detection of stable UTs so far {may still imply} extra mechanisms destabilizing Uranus companions. \citetads{2002Icar..160..271N} suggested that the significant depletion of the pre-formed population of primordial Trojans results from the MMRs in the radial migration of planets. Hence, we numerically studied the dynamical behaviour of pre-formed UTs during the early evolution of the outer solar system.

Even though the Nice model \citepads{2005Natur.435..466G,2005Natur.435..462M,2005Natur.435..459T} depicts a more realistic planetary migration process, we still refer to the radial migration in our simulations, where Jupiter migrates slightly inward while the other three giant planets migrate outward \citepads{1984Icar...58..109F,1996P&SS...44..431F,1999AJ....117.3041H,2002MNRAS.336..520Z}. This simplified model is appropriate for the purpose of investigating the survivability of primordial UTs in this paper and it requires only a limited computing resources. This process can be conveniently simulated via an artificially constructed force instead of the interaction with the planetesimal disk. An exponential migration is adopted so that the semi-major axes of the planets could evolve according to
		\begin{equation}
     		a_j(t)=a_j(t=\infty)-\delta{a_j}\,\exp(-t/\tau)\,,
   		\end{equation}
where $a_j(t)$ is the semi-major axis of the $j^{\rm th}$ planet after time $t$, $\delta{a_j}$ is the desired amount of the total radial migration and $\tau$ is the e-folding time determining the rate of migration \citepads{1984Icar...58..109F,1996P&SS...44..431F,1999AJ....117.3041H,2002MNRAS.336..520Z}. Following the initial planetary configuration and $\delta{a_j}$ ($-0.2$, 0.8, 3.0 and 7.0\,AU for four giant planets respectively) adopted in \citetads{2004Icar..167..347K}, we have reproduced the fast and slow migration in our simulations with $\tau=1$\,Myr and 10\,Myr, respectively.  
	
In view of that no significant asymmetry is found between $L_4$ and $L_5$ during the radial migration \citepads{2015Icar..247..112P}, we consider an initial swarm of massless planetesimals formed near Uranus and were trapped in the tadpole region around $L_4$ while Uranus was growing, before the planetary migration. We use the same strategy as previously described in Sect.~\ref{subsec:modinit} to produce the initial orbital elements of pre-formed UTs. Since the disk may be pre-heated before the final stage of planetary migration \citepads{2015Icar..247..112P}, we choose an initial inclination distribution of pre-formed UTs that is wide enough ($0^\circ$--$65^\circ$). After some test runs, totally 392 initial semi-major axes are evenly distributed from 16.060 to 16.451\,AU around the initial position of Uranus (16.316\,AU), and finally 51\,352 ($=392\times131$) pre-formed UTs are set. By checking the libration amplitude during the first 25\,000 years, we verify that most of the sampled orbits are bound to the 1:1 MMR with Uranus initially. For the fast and slow migration, there are 35\,691 and 36\,607 such orbits in total respectively and we use them as the original population. Then the whole system is integrated for $5\tau$, when the migration is 99.33\% complete.
	
\begin{figure*}
	\resizebox{\hsize}{!}{\includegraphics{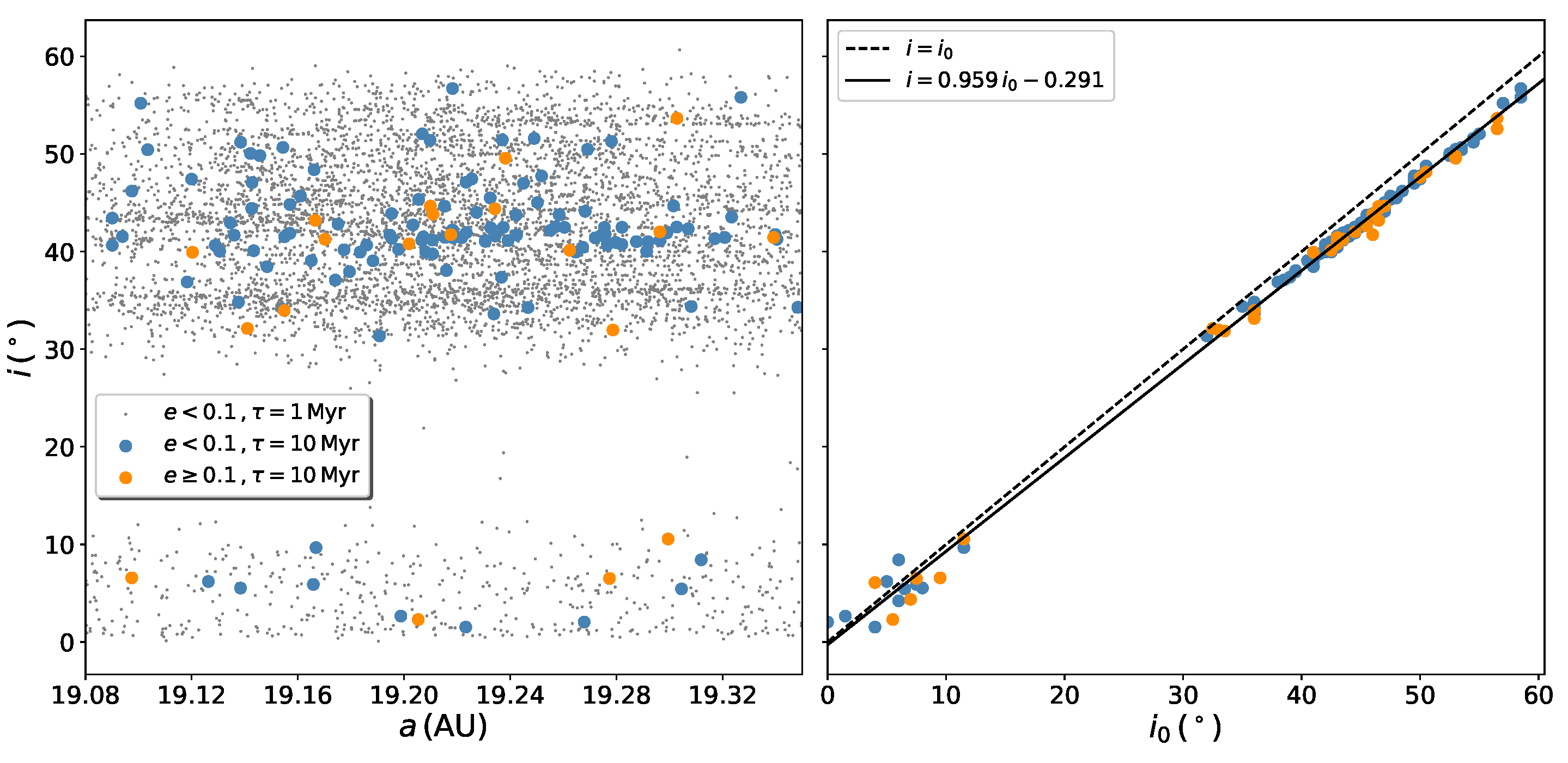}}
	\caption{Final distributions of orbits that have never left the 1:1 MMR region of Uranus during planetary migration. The \textit{left} and \textit{right panels} show the final inclination $i$ against the final semi-major axis $a$ and initial inclination $i_0$, respectively. The black dots indicate the orbits surviving the fast migration with the final eccentricity $e$ smaller than 0.1 while the others indicate the orbits surviving the slow migration, of which the blue and orange points represent those with $e<0.1$ and $e\ge 0.1$, respectively. The dashed line in the \textit{right panel} depicts the equality of $i$ and $i_0$ while the solid line depicts the best fit for the slow migration. For a clear comparison with the stability regions (see Figs.~\ref{fig:sn} and \ref{fig:lf}), we show the same range in semi-major axis in the \textit{left panel} at the cost of abandoning a small part of orbits. In the \textit{right panel} we only show the data for the slow migration for a better vision.}
	\label{fig:pltmig}
\end{figure*}

The final distributions of the orbits that have never left the co-orbital region are shown in Fig.~\ref{fig:pltmig}. Compared to the slow migration (0.4\%), much more orbits (36.3\%) survive the fast migration, apparently. At the same time, high-inclined orbits have a much greater chance of surviving the migrations. The same instability gap as in the dynamical maps induced by $\nu_5$ and $\nu_7$ (see Fig.~\ref{fig:gsres}), especially the later in this case, appears for both migration rates, occupying the moderate-inclination regime from $\sim10^\circ$ to $\sim30^\circ$. 

Recalling the stability regions as well as the orbits surviving the long-term evolution (Figs.~\ref{fig:sn} and \ref{fig:lf}), we estimate the probability of the orbits residing in the regions where they can survive the age of the solar system at the end of the migrations. The probability is obtained by calculating the product of the proportion of migratory orbits ending up in the regions enclosed by the stable orbits {(two rectangles on the $(a,i)$ plane that tightly cover the white points in Fig.~\ref{fig:lf}, one for high inclination and the other for low inclination)} and the proportion of area occupied by the stable orbits in those regions in current planetary configuration {(i.e. ratio of number of white points to the total number of points in the corresponding rectangle)}. The results are $6.69\times10^{-3}$ and $1.16\times10^{-4}$ for the fast and slow migrations respectively. As mentioned before, the eccentricity {of long-term} stable UTs should be smaller than 0.1 and under this constraint, the corresponding probabilities drop to $4.06\times10^{-3}$ and $9.07\times10^{-5}$, of which 77\% is contributed by the low-inclined orbits for both kinds of migrations. However, it should be noted that the final configuration of both the fast and slow migrations cannot exactly match the current one which is used to construct the dynamical maps.

Fig.~\ref{fig:pltmig} also presents the relation between the initial ($i_0$) and final inclinations ($i$) of orbits surviving the slow migration. In general, the surviving orbits end up with inclinations slightly smaller than their original values. We numerically fit the relation with a linear equation which is close to $i=i_0$. It is similar to the results from \citetads{2015Icar..247..112P}, where the inclinations change little for captured Neptune Trojans from the planetesimal disk. Actually, the similar relation also applies to the fast migration. If there remain any high-inclined UTs at the end of the migration, it comes from the original inclinations, instead of the stirring from the migration process as the migration does not heat the Trojans, {and even marginally cools them in the migration models adopted by us}. And in fact, the capture process also prefers the high-inclined orbits \citepads{2015Icar..247..112P}, just as the pre-formed UTs with high inclinations have a larger chance of surviving the migration.

\begin{figure}
	\resizebox{0.95\hsize}{!}{\includegraphics{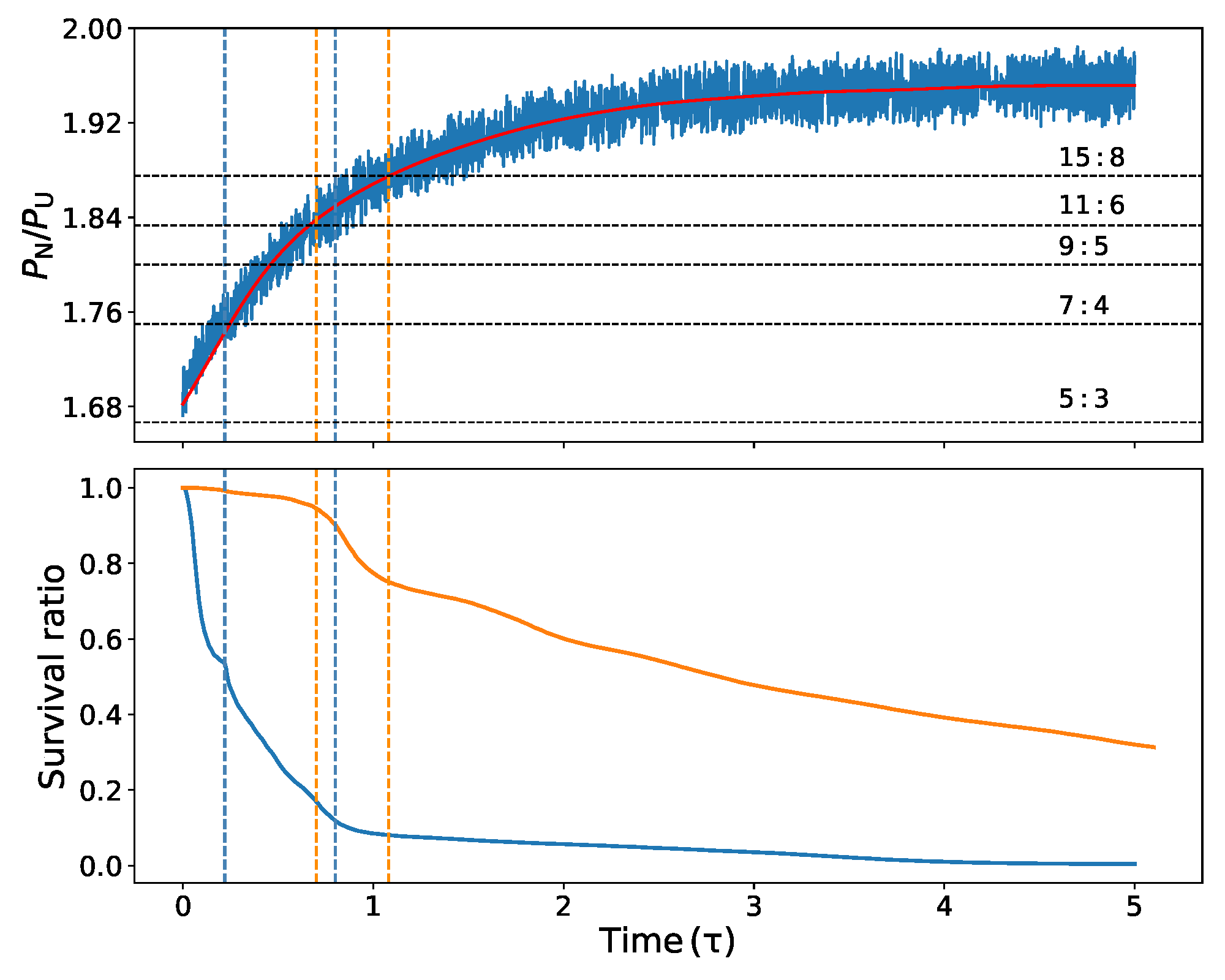}}
	\caption{Orbital period ratio of Neptune and Uranus ${P_{\rm N}/P_{\rm U}}$ (\textit{top panel}) and survival ratio of the original population (\textit{bottom panel}) during the fast (orange) and slow (blue) migrations. {The orbital period ratio of which the short-period terms are eliminated by the low-pass filter is indicated by the red line.} The horizontal lines indicate the locations of the 15:8, 11:6, 9:5, 7:4 and 5:3 MMRs between Neptune and Uranus. Two important moments for the fast migration, 0.70\,$\tau$ and 1.08\,$\tau$, and two important moments for the slow migration, 0.22\,$\tau$ and 0.80\,$\tau$, are indicated by the vertical lines. Note that we only show the period ratio for the slow migration in the \textit{top panel}, and the one for the fast migration is almost the same.}
	\label{fig:migmmrs}
\end{figure}

The decay curve of the Trojan population shown in Fig.~\ref{fig:migmmrs} illustrates that most pre-formed UTs are depleted in the first $10^7$\,yr of the slow migration. During this period, Neptune and Uranus have crossed their mutual MMRs, such as the 7:4, 9:5 and 11:6 MMRs. Concretely, every time the ice giants approach one of these MMRs, the secondary resonances associated with the MMR period and the libration period of UTs arise, inducing instabilities on the companions of Uranus \citepads{2004Icar..167..347K}. The evolution of the period ratio of Neptune and Uranus reveals the moments when the ice giants cross the important MMRs (Fig.~\ref{fig:migmmrs}). Obviously, the turning point of the decay curve at $0.22\,\tau$ results from the crossing of the 7:4 MMR between Neptune and Uranus. As these two planets get far from the 11:6 MMR from $0.80\,\tau$, the depletion rate of the UT population slumps. Since then, the approach of Neptune and Uranus to the 1:2 MMR and other weaker MMRs may have contributed to the loss of UTs. For the fast migration, the depletion rate is much smaller due to shorter durations in the secondary resonances except the period from $0.70\,\tau$ to $1.08\,\tau$, when the ice giants are moving from the 11:6 MMR to the 15:8 MMR.

\subsection{Chaotic capture}

As the dynamical evolution of a gravitating system is a time-reversible process, asteroids can enter the co-orbital region when the instability drives the Trojans out. Hence the instabilities induced by the secondary resonances during the migration also paves the way for the so-called chaotic capture, which means Trojans could be captured when Uranus and Neptune cross their mutual MMRs \citepads{2005Natur.435..462M,2009AJ....137.5003N,2010MNRAS.405.1375L}. Therefore, in our simulations, at the moments such as $0.22\,\tau$, $0.80\,\tau$ and $1.08\,\tau$, when the ice giants cross the 7:4, 11:6 and 15:8 MMRs (see Fig.~\ref{fig:migmmrs}), many objects in the disk would be captured as supplement to the co-orbital region of Uranus. Once the planets get far away from these MMRs, the co-orbital regions become temporarily stable until they approach the next MMR. In the end, there will always be some Trojans left, trapped in the co-orbital region. 
	
\begin{figure}
	\resizebox{0.95\hsize}{!}{\includegraphics{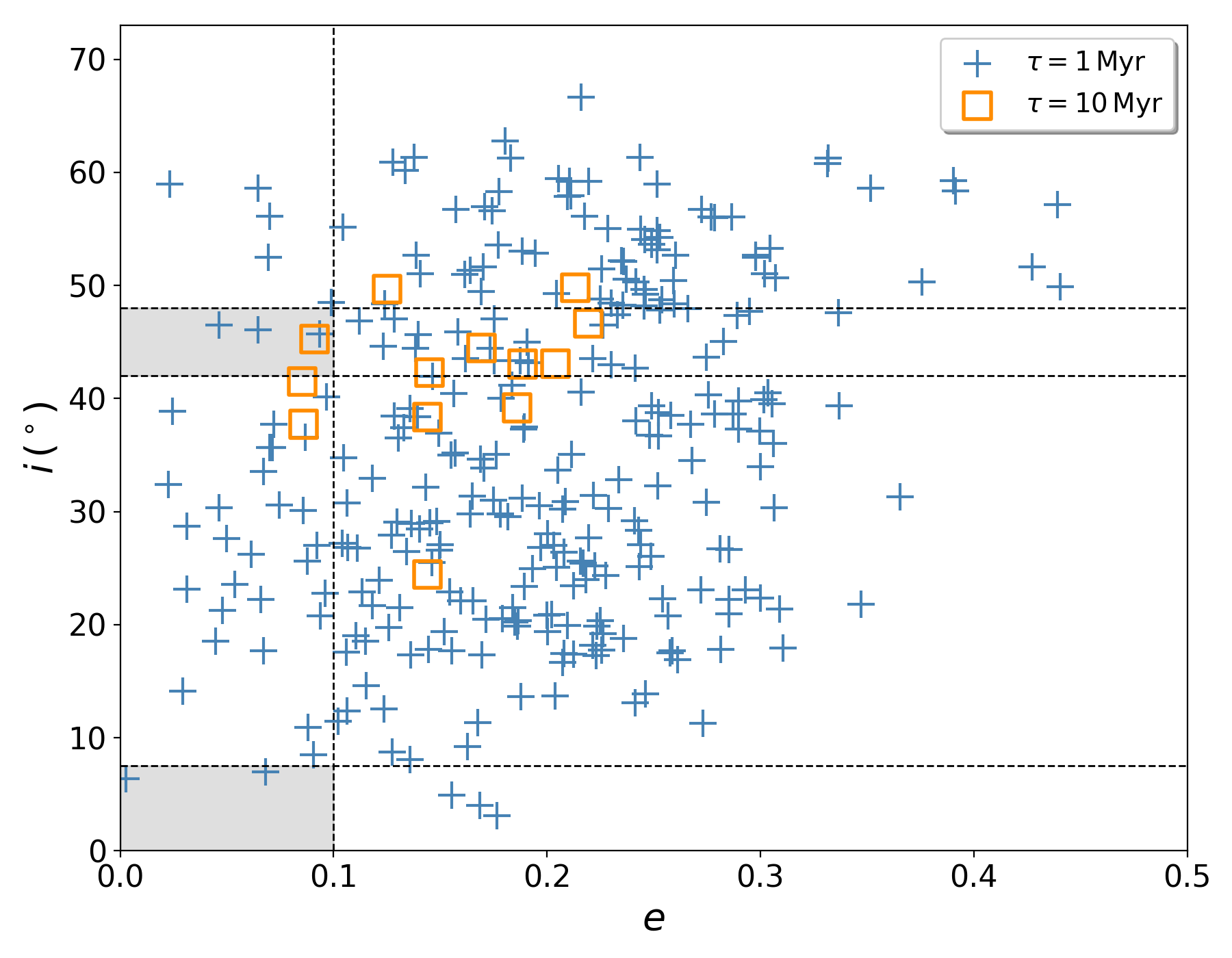}}
	\caption{Distribution of orbits recaptured as Uranus Trojans at the end of the simulations. The {blue} crosses indicate the fast migration ($\tau=1$\,Myr) while the {orange} squares indicate the slow migration ($\tau=10$\,Myr). The critical eccentricity for stable UTs $(e=0.1)$ is depicted by a vertical line. The horizontal lines indicate the critical inclinations for the long-life UTs, which are $7.5^\circ$, $42^\circ$ and $48^\circ$ {(see Fig.~\ref{fig:lf})}. The shaded areas indicate the stable regions.}
	\label{fig:recap}
\end{figure}

Besides those planetesimals shown in Fig.~\ref{fig:pltmig} that have never left the co-orbital region, much more objects have escaped during the migration. Many of them may enter into the co-orbital region again later, and such recapture must be comparable to the chaotic capture from the planetesimal disk. Many objects are found to have been recaptured at least once in our simulations. Of course, the depletion and recapture would compete with each other, adjusting the amount of the UT population. By the end of the migration, there are certain number of such recaptured objects in the co-orbital region, but here we only count those that stay in the co-orbital region for more than 1\,Myr after being recaptured. The orbital distribution of the final swarm of captured UTs is shown in Fig.~\ref{fig:recap}. Again, more orbits can be recaptured during the fast migration. Compared to the pre-formed UTs surviving the migrations, much fewer orbits end up being recaptured. Almost none of these orbits falls in the stability region in the low-inclination regime ($i<7.5^\circ$). The other stability region for the high-inclined UTs hold most of the recaptured orbits in the slow migration. However, only one of them has a little chance to survive the age of the solar system ($42^\circ<i<48^\circ$, $e<0.1$). In consideration of the low survival probability for high-inclined UTs, we speculate that it is nearly impossible to detect the recaptured orbits nowadays. 

In view of the large amount and different orbital distribution of the planetesimals, the realistic chaotic capture from the disk could be more efficient. \citetads{2010MNRAS.405.1375L} investigated the capture of Trojans from the planetesimal disk by Uranus with different migration rates and different disk models. A capture efficiency between $10^{-5}$ and $10^{-4}$ is obtained. Although the capture efficiency is sensitive to the model of the planetesimal disk, it is still plausible that the chaotic capture could produce some survivors. Nevertheless, the orbital distribution of the captured UTs  also suggests that their orbits are kind of hot and they are unlikely to survive the age of the solar system \citepads{2010MNRAS.405.1375L}. 
	
Although there is no definitive evidence yet, some studies have shown the preference for the slow migration with $\tau\sim 10$\,Myr. High inclinations of plutinos support a migration time scale about 10\,Myr \citepads{1998LPI....29.1476M}. In addition, \citetads{2009MNRAS.398.1715L} demonstrated that the slow migration where Neptune migrates over a large distance (from 18.1\,AU to its current location) could produce an orbital distribution of Neptune Trojans more matching the known one than the fast migration. {The process how the planets migrate during the early stage of the solar system only determines the original population of UTs rather than the stability in the current planetary configuration.}
	
The Nice model suggests that Jupiter and Saturn have crossed the 1:2 MMR before the smooth migration that can be approximated by our migration model, but with possibly different time scales $\tau$ \citepads{2016A&A...592A.146G}. The 1:2 MMR crossing of Jupiter and Saturn would cause global instability, which could empty the pre-formed UTs due to a violent change of the orbit of Uranus. This further reduces the possibility of observing primordial UTs nowadays. But in turn, {according to the well-adopted migration models now}, if any primordial UTs are observed in future, it will suggest that there may be a swarm of pre-formed UTs at the beginning of the smooth outward migration of Uranus, which is more likely of a short migration time scale.
	
%-----------------------------------------------------------------

\section{Conclusion and Discussion}\label{sec:disc}

Uranus is one of the five planets in the solar system which are verified to hold Trojan asteroids. However, confirmed UTs and candidates detected so far are all temporarily captured. In this paper, we investigated the stability of UTs via numerical simulations. Detailed dynamical maps were constructed on the $(a_0,i_0)$ plane with the SN as the stability indicator. Besides a stability region at low inclination ($0^\circ$--$14^\circ$), stable UTs can also reside in the stability window ($32^\circ$--$59^\circ$) in the high-inclination regime {(see Fig.~\ref{fig:sn})}. Two instability gaps appear around $9^\circ$ and $51^\circ$, dividing both stability regions into two. Besides, abundant fine structures are present in the dynamical maps. These instability gaps block the possible route of Trojans achieving high inclination through dynamical diffusion in the phase space. 
	
In order to locate the resonances shaping the structure of the phase space, we implemented a frequency analysis method to portray the resonance web on the $(a_0,i_0)$ plane. The apsidal secular resonances with Jupiter ($\nu_5$) and Uranus ($\nu_7$), located at moderate inclination, are the strongest mechanisms destabilizing UTs {(see Fig.~\ref{fig:gsres})}. These secular resonances could significantly excite the eccentricity of orbits, %based on their distance from these resonances, 
and help clean the moderate-inclined UTs. Two instability gaps around $9^\circ$ and $51^\circ$ are believed to be involved with the apsidal resonances $g-2g_5+g_7=0$ and $\nu_8$. High-degree secular resonances and secondary resonances could shape the outline of the stability region and carve the fine structures.
	
The orbits are mainly excited by secular resonances, in both eccentricity and inclination. Combined secular resonances such as $g+2s-s_5-g_8-s_8=0$ could pump up the inclination with an amplitude up to $\sim 12^\circ$. This will not form high-inclined UT population because the involved orbits are not stable, but it may contribute to the excitation of Centaurs' inclination since they generally spend portion of lifespan on the temporary Trojan orbits \citepads{2013Sci...341..994A}.

Kozai mechanism plays an important role in controlling the variation of eccentricity and inclination. It is also worth noting that UTs will experience close encounters with the parent planet when they are on large-amplitude horseshoe or quasi-satellite orbits, leading to a huge increase in eccentricity {(see Fig.~\ref{fig:obtexmp})}. Actually, UTs with eccentricity larger than 0.1 will lose their stability sooner or later.
	
Although some high-inclined UTs within the stability region have small libration amplitudes of the resonant angle, the most stable UTs at low inclination are of relatively larger libration amplitudes {(see Fig.~\ref{fig:dsigma})}. All orbits that have ever left the tadpole cloud around $L_4$ have a libration amplitude larger than $300^\circ$. In fact, they keep switching between different co-orbital states after leaving the tadpole orbit and we confirm that there lack stable horseshoe orbits for UTs {(see Fig.~\ref{fig:obtexmp})}.
	
The libration center where the libration amplitude is the smallest varies with the initial inclination {(see Fig.~\ref{fig:dsigma})}. It is related to the quasi 1:2 MMR between Uranus and Neptune as the displacement will disappear if Uranus and Neptune were further away from the 1:2 MMR. However, the secondary resonances involving the frequency of the quasi 1:2 MMR seem to have no effect on the long-term stability of UTs {(see Fig.~\ref{fig:fres})}.
	
A long-time simulation up to the age of the solar system (4.5\,Gyr) reveals the residence of possible primordial UTs. They are close to the libration center {(see Fig.~\ref{fig:lf})}. About 3.81\% of UTs can survive the integration, of which 95.5\% are on low-inclined orbit. Another 4.5\% orbits are distributed in the range of $i_0=42^\circ$ to $48^\circ$. The sky densities show that the primordial UTs are most likely to be detected on coplanar orbit of Uranus, but for high-inclined ones, the region where the relative longitude and latitude with respect to the instantaneous orbital plane of Uranus are $72^\circ$ and $\pm40^\circ$ could be an alternative {(see Fig.~\ref{fig:sky})}.
	
In spite of a nonzero survival ratio, no primordial UTs have been found in observation so far. \citetads{2017MNRAS.467.1561D} demonstrated that ephemeral multi-body mean motion resonances may lead to the depletion of primordial UTs in the early evolution of the solar system. In our study, we consider the radial migration which could destabilize the primordial UTs before the formation of the current planetary configuration. 
	
The planetary migration with time scales $\tau=1$ and 10\,Myr is adopted. For the fast migration, 36.3\% of the pre-formed orbits survive the migration to $5\,\tau$ while it is only 0.4\% for the slow migration {(see Fig.~\ref{fig:pltmig})}. In general, the surviving orbits are of inclinations slightly smaller than their initial values, implying that the migration process cannot excite the inclinations. In spite of a relatively small amount, low-inclined orbits still contribute 77\% of the orbits that can survive the age of the solar system after the migration. These stable orbits occupy $4.06\times10^{-3}$ and $9.07\times10^{-5}$ of the origin population for the fast and slow migrations, respectively.
	
The depletion of the pre-formed UTs result from the MMR crossing of Uranus and Neptune. Once they approach the mutual MMR, the related secondary resonances would destabilize the UTs {(see Fig.~\ref{fig:migmmrs})}. Nevertheless, the orbits outside the co-orbital region can be captured at the same time. %as the dynamical evolution is time-reversible. 
The orbital distributions from both our simulations %using the escaped orbits 
and the chaotic capture of the planetesimals \citepads{2010MNRAS.405.1375L} support that the captured orbits are unlikely to survive to today {(see Fig.~\ref{fig:recap})}.

Actually, the final state of the migration depends on the parameters of the migration model, including the migration rate and the initial planetary configuration, especially the locations of Uranus and Neptune. It is suggested that increasing migration time scale ($\tau$) could reduce the survival ratio over the planetary migration \citepads{2004Icar..167..347K}. \citetads{2014Icar..232...81M} demonstrated that Neptune may have undergone the migration with a much longer migration time scale ($\sim100$ Myr). Some simulations of the radial migrations also include the eccentricity damping since a large eccentricity of Neptune may be necessary at the beginning of the smooth outward migration \citepads{2012ApJ...750...43D,2012ApJ...746..171W,2015Icar..247..112P}.
	
Compared to the radial migration we use, the Nice model where the migration is a natural result of the gravitational interaction between the planets and planetesimals is more realistic and promising. Its most significant difference from our model is the instability phase when Jupiter and Saturn crossed the mutual 1:2 MMR before the smooth migration. This MMR crossing could make a great change to the orbits of Uranus and Neptune, causing a severe depletion of their pre-formed Trojans. The stirring of the planets owing to the instability phase can be relieved afterwards by the dynamical friction from the planetesimal disk \citepads{2005Natur.435..459T}. But in fact, our model could approximate the migration after the instability phase in the Nice model and the mechanisms of the depletion and chaotic capture then are nothing different \citepads{2009MNRAS.398.1715L,2009AJ....137.5003N,2010MNRAS.405.1375L,2016A&A...592A.146G}.
	
In the Nice model, Neptune may have been further {away} from the Sun before the end of the migration, which means Uranus could be closer to the 1:2 MMR with Neptune than it is today \citepads{2016A&A...592A.146G}. We investigate the orbital behaviour of UTs with a further location of Neptune and find that the 1:2 MMR between Uranus and Neptune greatly destabilize the UTs, driving most of them out of the co-orbital region in a short time. In other words, if Uranus and Neptune have been closer to the 1:2 MMR in the past, we can hardly observe any primordial UTs today.
	
%-----------------------------------------------------------------

\begin{acknowledgements}
      Our sincere appreciations go to the anonymous referee, whose comments are very helpful in improving this paper. This work has been supported by the National Natural Science Foundation of China (NSFC, Grants No.11473016, No.11933001 \& No.11973027). R. Dvorak wants to acknowledge the support from the FWF project S11607/N16.
\end{acknowledgements}


%\begin{thebibliography}{}

 % \bibitem[Murray \& Dermott(1999)]{1999ssd..book.....M} Murray, C.~D., \& Dermott, S.~F.\ 1999, solar system dynamics by C.D.~Murray and S.F.~McDermott.~(Cambridge, UK: Cambridge University Press),  ISBN 0-521-57295-9 (hc.), ISBN 0-521-57297-4 (pbk.).
  \bibliographystyle{aa-note} % style aa.bst
  \bibliography{UTref}
   
%\end{thebibliography}

\end{document}